\documentclass[aps,pra,reprint]{revtex4-2}

\usepackage{graphicx,mathtools,epstopdf,amssymb,amsmath,commath,verbatim,csquotes,hyperref,ulem}
\usepackage{pifont}
\usepackage{tikz}
\newcommand*\circled[1]{\raisebox{.5pt}{\textcircled{\raisebox{-.9pt} {#1}}}}
        
\DeclareMathOperator{\sech}{sech}
\DeclareMathOperator{\csch}{csch}
\usepackage{color}

\begin{document}
	\title{Collision dynamics of discrete soliton in uniform waveguide arrays}
	\author{Anuj P. Lara}
	\affiliation{Department of Physics, Indian Institue of Technology Kharagpur, Kharagpur 721302, India}
	\author{Samudra Roy$^{\dagger}$}
	\affiliation{Department of Physics, Indian Institue of Technology Kharagpur, Kharagpur 721302, India}
	\email{$^\dagger$samudra.roy@phy.iitkgp.ac.in}
	
	\begin{abstract}
		We investigate the collision dynamics of discrete soliton (DS) pair in a realistic semi-infinite nonlinear waveguide array (WA) in the context of \textit{diffractive resonant radiation} (DifRR). Depending on the initial amplitude ($A_0$) and wavenumber ($k_0$), a co-moving pair of identical DSs either collide elastically  or merge to form a discrete \textit{breather}.  For large amplitude and small wavenumber, DSs form a bound state and do not interact .  
		We map the domain of interaction by iterative simulation and present a phase plot  in ($A_0$-$k_0$) parameter space.
		A variational technique is developed where the interaction term is considered as perturbation. A proper choice of Lagrangian density and the \textit{ans\"{a}tz} function followed by the Ritz optimization leads us to a set of ordinary differential equations that describe the collision mechanism.  The analytical result corroborate well with numerical data.  
		We further investigate the role of initial phase detuning between two identical DS and observe a periodic energy exchange during their interaction. 
		Varied degree of energy exchange occurs when two DS with  different wavenumbers collide which results in three distinct output states accompanied by DifRR generation. 
		Extending our investigation to a more generalized condition by taking different amplitudes and wavenumber of DS pair, we find an unique secondary radiation in $k$-space which is originated due to the collision of solitons. 
		The nature of this collision mediated secondary radiation is found to be different from usual DifRR. 
		Our results shed light on the interesting aspects of the collision dynamics of DS pair in nonlinear WA and useful in understanding the complex mechanism.

	\end{abstract}
	\maketitle
	
	\section{Introduction}
	Optical waveguide arrays (WAs) have garnered considerable interest over the decades as a system for exploring discrete physical phenomena which are on the atomic scale and otherwise difficult to study in a laboratory environment, and as photonic components for integrated optical devices.
	These arrays composed of evanescently couple waveguides have been modeled by the introduction of  coupled mode theory for two adjacent waveguides\cite{millerCoupledWaveTheory1954,yarivCoupledmodeTheoryGuidedwave1973,dasNonlinearEffectsCoplanar1987} and the extension thereafter to multiple waveguide systems\cite{hausCoupledMultipleWaveguide1983,finlaysonSpatialSwitchingInstabilities1990}.
	In the continuous approximation, beam propagating in nonlinear WA obeys the nonlinear Schr\"{o}dinger equation and exhibit self focusing and localization \cite{christodoulidesDiscreteSelffocusingNonlinear1988a,sipeNonlinearSchrodingerSolitons1988a,kivsharPeierlsNabarroPotentialBarrier1993,kivsharSelflocalizationArraysDefocusing1993}.
	Optical Kerr nonlinearity, in balance with discrete diffraction, an analogue of diffraction in continuous media \cite{pertschAnomalousRefractionDiffraction2002a,eisenbergDiscreteSpatialOptical1998} exhibits optical discrete solitons (DSs), which are observed both theoretically and experimentally\cite{eisenbergOpticalDiscreteSolitons2002a,OpticalDiscreteSolitons2002,acevesDiscreteSelftrappingSoliton1996,caiLocalizedStatesDiscrete1994}.
	These spatial solitons however manifest properties that are intriguing and forbidden in the case of their continuous counterpart.
	The discrete nature of these WAs and the periodic potential that is formed thereby, facilitates the study of many fundamental phenomena present in discrete systems like atomic and molecular lattices on a macroscopic scale.
	These WAs exhibit properties and phenomena like Peierls'-Nabbaro potential\cite{morandottiDynamicsDiscreteSolitons1999}, anomalous refraction and diffraction with limitation in transverse energy transportation\cite{pertschAnomalousRefractionDiffraction2002a}, Anderson localization\cite{lahiniAndersonLocalizationNonlinearity2008,martinAndersonLocalizationOptical2011} Bloch Oscillations, localized Wannier-Stark States \cite{morandottiExperimentalObservationLinear1999,pertschOpticalBlochOscillations1999}, Bloch-Zener Oscillation and Zener tunneling\cite{breidBlochZenerOscillations2006,dreisowBlochZenerOscillationsBinary2009}.
			With rapid progress in fabrication techniques, specific WAs with novel structures can be designed. For example, sinusoidally curved WAs \cite{longhiObservationDynamicLocalization2006}, interlaced two component super-lattice as a base for binary WAs \cite{hizanidisInterlacedLinearnonlinearOptical2008},  two-dimensional WAs  for continuous-discrete systems supporting spatiotemporal solitons (3D Light Bullets) \cite{minardiThreeDimensionalLightBullets2010}, dissipative as well Ginzburg-Landau solitons, and surface solitons\cite{mihalacheSpatiotemporalDissipativeSolitons2008a,mihalacheSpatiotemporalSurfaceGinzburgLandau2008}.
	Binary WAs further offer an optical approach to study relativistic phenomenon such as \textit{Zitterbewegung}\cite{longhiPhotonicAnalogZitterbewegung2010}, Klien tunneling\cite{longhiKleinTunnelingBinary2010}, Fock states\cite{keilClassicalAnalogueDisplaced2011}, neutrino oscillations\cite{mariniOpticalSimulationNeutrino2014}, and Dirac solitons \cite{tranOpticalAnalogueRelativistic2014} for instance.
	Additionally, controlling beam and pulse propagation in light bullet routing\cite{williamsGeneratingRoutingLightbullets2012a}, discrete soliton routing by external fields\cite{zhangManipulatingDiscreteSolitons2017} have been demonstrated and studied.
	Further progress in the recent years have been extended to the regime of plasmonics with discrete diffraction and Bloch oscillations in plasmonic waveguide arrays\cite{blockBlochOscillationsPlasmonic2014,pezziPlasmonmediatedDiscreteDiffraction2019}.
	With such a wide range of studies, the versatility of WA systems have been proven and may provide as a base for further work in the upcoming future.
		
		In the simplest WA structure where the separation of the adjacent waveguide channel is constant, the evolution of modes in individual waveguides that are coupled with their nearest neighbor waveguides is governed by the standard coupled mode equations.
	These equations also take into account the linear and nonlinear terms and the set of equations combined together form what is known as the discrete nonlinear Schr\"{o}diner equation (DNLSE). In such WAs, also referred to as homogeneous WAs, DSs are formed as a result of the balance between discrete diffraction and self focusing Kerr nonlinearity.
	Different properties of DSs have been studied over the decades since they were first observed; however the phenomenon of emission of radiation from these solitons is a recent development in comparison to all the previously stated works.
	This radiation, emitted by a spatial DS propagating with a transverse velocity component in an uniform WA, is aptly named discrete diffractive resonant radiation (DifRR) \cite{tranDiffractiveResonantRadiation2013}.
	 The formation of DifRR under a phase matching (PM) condition is analogous to the \textit{dispersive wave} (DW) generation in its temporal counterpart \cite{KarpmanPRE}.
	DW  emerges owing to the perturbation of temporal solitons by higher-order dispersion and nonlinearity of the optical fiber. 
	The spectral location of DW radiation is sensitive to the zero-dispersion frequency and can be tuned to excite from deep UV to far IR regime \cite{Roy:09}.  
	However, in WA, due to the one-dimensional (1D) lattice formed by the periodic arrangement of the waveguides, the special DS as well as the generated DifRR exist within the resulting Brillouin boundary and their wavenumbers are limited to $-\pi$ and $\pi$.
	This results in $2\pi$ shift in any electric field that crosses the Brillouin boundary and recoil from the opposite side yielding \textit{anomalous recoil} \cite{tranDiffractiveResonantRadiation2013}.
	An initial phase gradient or wavevetor is required to \textquote{push} the DS away from a normal incidence and interact with the waveguide with a transverse component attributing exciting phenomenon like DifRR generation.
	It is only natural to be curious about how two of these special DSs with opposite wavenumbers (or \textit{push}) will interact with each other in the lattice like structure of the WA.
	However, very limited background studies regarding its dynamics are presented which have been mostly limited to continuous spatial solitons\cite{krolikowskiFusionBirthSpatial1997,aosseyPropertiesSolitonsolitonCollisions1992,malomedPotentialInteractionTwo1998}, and a coupled set of NLS that form vector solitons \cite{anastassiouEnergyExchangeInteractionsColliding1999,anastassiouEnergyExchangeInteractionsColliding1999,vahalaInelasticVectorSoliton2004,katsimigaDarkbrightSolitonPairs2018,stalinNondegenerateBrightSolitons2021}
	Of the few mechanisms and phenomenon studied regarding DS are symmetry breaking and momentum nonconservation, velocity dependent soliton merger and breather formation\cite{papacharalampousSolitonCollisionsDiscrete2003,alkhawajaInteractionPotentialDiscrete2016}, soliton collision in optically induced photonic lattice\cite{xiaoTunableOscillationDiscrete2011} and  WAs with saturable nonlinearity\cite{cuevasDiscreteSolitonCollisions2006}.
	Studies on structures that are governed by DNLSE, for example a classic ferromagnetic spin chain with \textit{Dzyaloshinskii-Moriya} interaction \cite{parasuramanDynamicsSolitonCollision2019a} has also been shown to have elastic soliton-soliton interaction akin to their continuous counterparts.
	Additional discrete system in which similar soliton-solition interaction have been studied are in Bose-Einstein condensates (BECs) \cite{gaoBreathingSolitonsInduced2021a,katsimigaDarkbrightSolitonPairs2018}.

	The primary focus of this study is to investigate the collision mechanism of a pair of identical and nonidentical DSs while considering the recent works on emission of DifRR. We propose a realistic WA that supports DS generation and the formation of DifRR. The most basic result of the interaction between two solitons is their elastic collision.
	For solitons in coupled systems, namely vector solitons, it has been observed to result in inelastic collision \cite{christovInelasticitySolitonCollisions1994}, and transfer of energy between them \cite{soljacicCollisionsTwoSolitons2003,vahalaInelasticVectorSoliton2004}.
	Intuitively, similar behavior can be expected for DSs and the same has been observed in the case of Kerr \cite{papacharalampousSolitonCollisionsDiscrete2003} and saturable \cite{cuevasDiscreteSolitonCollisions2006} nonlinearity, where solitons  are merged to form a \textit{breather}.
	It is to be expected, DSs to undergo elastic collision in the WA system  without any interactions other than a phase shift similar to their continuous counterparts.
	Based on this knowledge we perform an initial analysis of identical DSs colliding in a realistic WA.
	Of the many parameters of DSs in WAs, we find that the soliton wavenumber ($k_0$) and amplitude ($A_0$) determine whether the interaction yields elastic collision or a \textit{fused state} in the form of \textit{breather} formation. It is also observed that for higher $A_0$ and lower $k_0$ soliton collision is prohibited and we may obtain a bound state formed by DS pair. 
	Performing iterative simulation we develop a phase diagram in parametric space which determines the set of values ($k_0$,$A_0$) for which elastic collision, \textit{breather} or bound-state formation takes place. 
	We highlight interesting features like significant variation of the \textit{breather} period ($\xi_p$) at phase boundary and a sudden phase shift of DS at the point of collision. When the soliton is not strongly localized to a single channel rather distributed covering few waveguides, the interaction behavior resembles closely to that of solitons in continuous media and can be analyzed  semi-analytically using \textit{variational method} \cite{PhysRevA.27.3135} where interaction term is taken as a perturbation. 
	Considering the transverse variable of the  propagation equation to be continuous we chose appropriate Lagrangian
	density and reduce it using suitable \textit{ans\"{a}tz} function. The reduced variational problem, followed by the \textit{Ritz optimization}, leads
	to a set of coupled ordinary differential equations (ODEs) that governs the evolution of individual soliton parameters under collision. Solving the ODE we trackdown important information like, the trajectory of DS, its wavenumber and phase evolution during collision. The variational results corroborate well with full simulation. 
	The role of initial phase detuning ($\Delta \Phi$) which determines the interaction potential between DS pair is also investigated. It is found that for $\Delta \Phi =\pi$  the DS interaction is always repulsive irrespective of the set of values ($k_0$,$A_0$), $i.e.$ \textit{breather} formation is prohibited for out of phase soliton pair ($\Delta \Phi =\pi$). Significant energy exchange is observed between DSs when $0<\Delta \Phi<\pi$. We scan the interaction dynamics near phase boundary by detuning  $\Delta \Phi$ in the range  $\Delta \Phi \rightarrow 0-2\pi$ and observed periodic energy exchange between DSs accompanied by the generation of weak DifRR due to collision. 
	The collision dynamics of two DSs with same amplitudes and different wavenumbers differs noticeably  from that of two identical DSs (having same amplitude and wavenumber).
	A transfer of energy between solitons occur in this case akin to that of vector solitons \cite{soljacicCollisionsTwoSolitons2003}.
	In addition to this, a soliton with higher amplitude and initial wavenumber radiates DifRR in Fourier-space.
	Due to DifRR being a relatively recent development in this field, we have limited understanding of the interaction dynamics of the two DSs in the context of emitted radiation.
	Taking these facts into account, we focus our investigation on the most general case where two non-identical DSs (with different wavenumbers and amplitudes) interacts.
	Interestingly, the collision of two non-identical DSs yields a secondary radiation different from the usual DifRR generated by either of the DS.
	However, the lack of momentum conservation in this interaction \cite{papacharalampousSolitonCollisionsDiscrete2003} pose a challenge in predicting and determining the properties of this additional radiation.
	Despite this limitation, we perform extensive simulations to analyze the collision dynamics and some new insight is gained regarding the generation of this secondary radiation.

	\section{Theory}
	An infinite array consists of identical and lossless waveguides is considered to be an ideal WA.
	However for practical feasibility, we chose a semi-infinite array with a large number  of waveguide channels that avoid any form of boundary interaction phenomenon. 
	The mode evolution in the $n^{th}$ waveguide for continuous-wave excitation under nearest-neighbor evanescent coupling is governed by the standard DNSLE\cite{kivsharPeierlsNabarroPotentialBarrier1993,kivsharSelflocalizationArraysDefocusing1993,christodoulidesDiscreteSelffocusingNonlinear1988a}:
	\begin{align}
		i \frac{d E_n}{dz} + C_{(n)}^{({n+1})}E_{n+1} + C_{(n)}^{({n-1})}E_{n-1} + \gamma_n\abs{E_n}^2 E_n = 0,
		\label{eq:DNLSE}
	\end{align}

\begin{figure}[h]
		\begin{center}
			\includegraphics[width=1.0\linewidth]{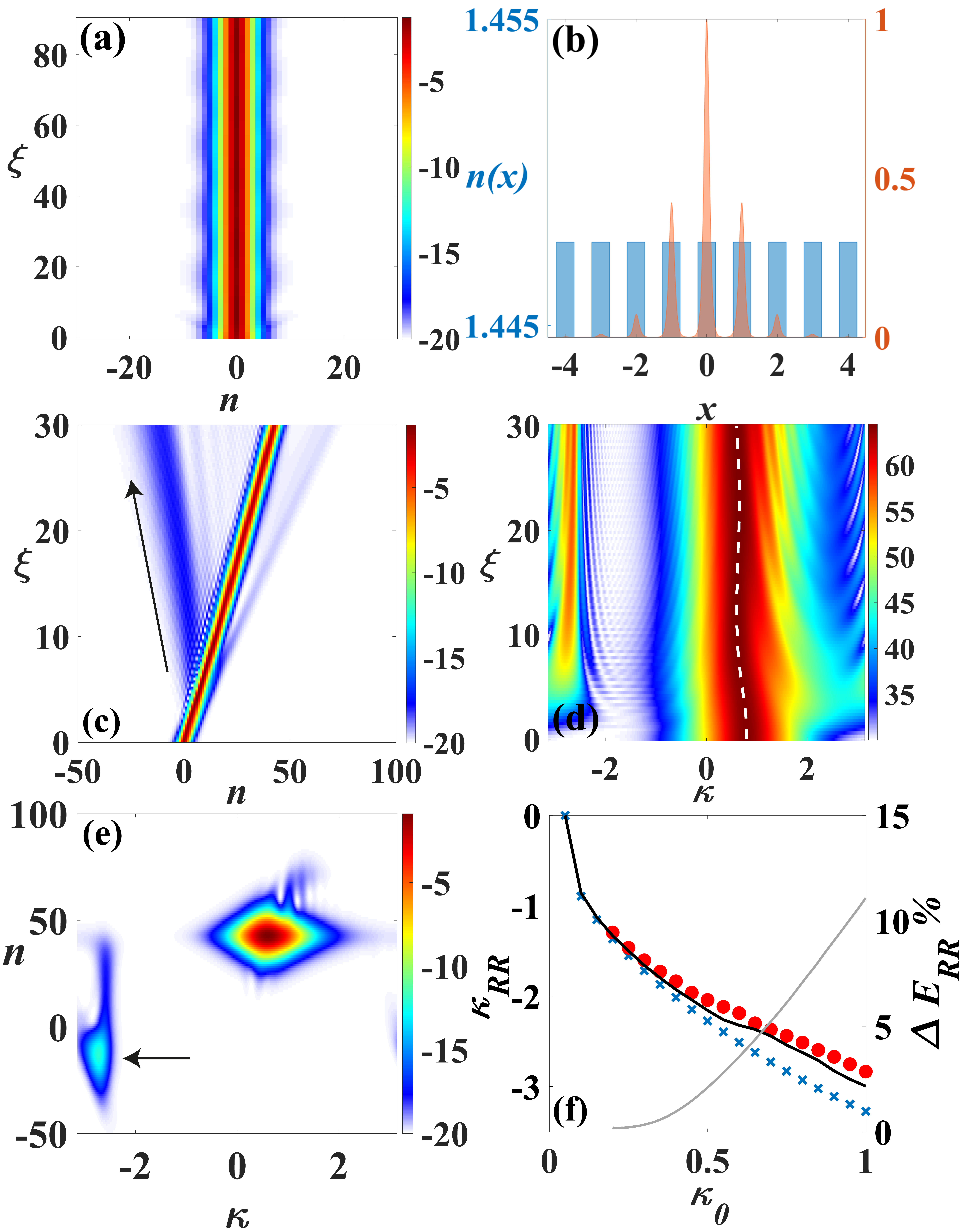}
			\caption{(a) Formation of DS in $n$ space propagating in a uniform WA for a soliton solution input. 
				(b) Spatial distribution of DS at output in the background of periodic refractive index profile $n(x)$ offered by the uniform WA.
				(c) Evolution of a DS launched at an incident wavenumber $\kappa_0 = 0.8$, emitting a diffractive radiation (direction marked by an arrow) opposite to the propagation of the DS.
				(d) The corresponding evolution in the $\kappa$-space with a strong DifRR emitted with wavenumber $\kappa_{RR} \approx -2.8$. The variation of the soliton wavenumber (soliton peak) is tracked by a white dotted line. 
				(e) Spectrogram depicting the location of DifRR (by arrow) at the output.
				(f) Location of generated DifRR as a function of incident angle $\kappa_0$ obtained numerically (red circles), from the PMC (crosses) and modified input for PMC (solid black line) using the average wavenumber.
				The solid gray line represents the fraction of energy carried by the DifRR ($\Delta E_{RR}$)  as a function of $\kappa_0$.
				In all the heat maps (a),(c),(d) and (e), along $z$-axis , $|\psi_n|^2$ and $|\tilde{\psi}_{\kappa}|^2$ are plotted in log scale for better resolution.}
			\label{fig:sol_lattice}
		\end{center}
	\end{figure}	
	\noindent where $E_n$ is the electric-field amplitude of the modes in the  $n^{th}$ waveguide.
	The range of the index $n$ is defined within $-N \leqslant n \leqslant N$, thereby defining the total number of waveguides to be $(2N + 1)$.
	$C_{(n)}^{(n+1)}$ and $C_{(n)}^{(n-1)}$ are the coupling coefficients of the $(n+1)^{th}$ and $(n-1)^{th}$ waveguide to the $n^{th}$ waveguide respectively with units in m$^{-1}$.
	$\gamma_n = \omega_0n_2/(cA_{eff})$ is the nonlinear coefficient of an individual waveguide in units of W$^{-1}$m$^{-1}$ where $n_2$ is the Kerr coefficient and $A_{eff}$ is the effective mode area.
	For a uniform WA the coupling coefficients are considered to be identical 
	 $\left(C_{(n)}^{(n+1)} = C_{(n)}^{(n-1)} = C\right)$. 
	Additionally, the nonlinear coefficient ($\gamma_n$) is equal for every waveguide and written as $\gamma_n = \gamma$ $\forall n$, as each of the constituent waveguides are considered to be composed of the same material and have the same dimensions.

	For low power condition the nonlinear term can be neglected $(\gamma = 0)$ and Eq.(\ref{eq:DNLSE}) thereby reduces to an analytically integrable equation.
	The solution of single waveguide excitation in such a case leads to mode evolution that exhibit discrete diffraction with the solution in the form of $E_n(z) = E_n(0)i^n J_n(2Cz)$, where $J_n$ is the Bessel function of order $n$\cite{eisenbergDiscreteSpatialOptical1998}.
	Physically, the varying $z$ dependent phase shift is the underlying reason for discrete diffraction.
	Intuitively, one can understand the formation of solitons by counteracting this discrete diffraction by a balancing Kerr induced self focusing\cite{christodoulidesDiscreteSelffocusingNonlinear1988a}.
	The DNLSE can be converted to an useful normalized form through the following transformations $E_n \rightarrow \sqrt{P_0 \psi_n}$, $\gamma P_0 z \rightarrow \xi$, and $C/(\gamma P_0)\rightarrow c$:
	\begin{align}
		i \frac{d \psi_n}{d \xi} + c\left[\psi_{n+1} + \psi_{n-1}\right] + \abs{\psi_n}^2 \psi_n = 0,
		\label{eq:nDNLSE}
	\end{align}
	where $P_0$ is the associated peak beam power in units of Watt.
	It is to be noted that the total power flowing through the array, $P = \sum_{n}\abs{\psi_n}$ and Hamiltonian, $H = \sum_n\left[c\abs{\psi_n - \psi_{n-1}}^2 - \gamma\abs{\psi_n}^2/2\right]$ remain conserved during the propagation\cite{morandottiDynamicsDiscreteSolitons1999} under idealized scenario (no losses and continuous wave excitation).
	In the linear case, exploiting the discrete plane-wave solution $\psi_n(\xi) = \psi_0\left[i(nk_xd + \beta \xi)\right]$ of Eq.\ref{eq:nDNLSE}, one can obtain the standard dispersion relation between $\beta$ and $k_x$ as	$\beta(\kappa) = 2c\cos(\kappa) + \abs{\psi_0}^2,$
		where $d$ is the separation between two adjacent waveguides, $k_x$ is the transverse wave vector and $\kappa \equiv k_x d$ is the phase difference between two adjacent waveguides\cite{christodoulidesDiscreteSelffocusingNonlinear1988a}.
	The transverse component ($\kappa$) undergoes a phase gain $\phi_t = \beta(\kappa)\xi$ during it's propagation which leads to the transverse shift of the propagating beam $\Delta n = \partial \phi_t/\partial \kappa$\cite{ledererDiscreteSolitons2002}.
	Hence, the beam propagates at an angle $\theta = \tan^{-1}\left[\partial \beta (\kappa)/\partial \kappa\right] = \tan^{-1}\left[-2c \sin(\kappa)\right]$\cite{eisenbergDiffractionManagement2000}.
	The Taylor expansion of $\beta(\kappa)$ about the incident wavenumber ($\kappa_0$) results in the diffraction relation:
	\begin{align}
		\beta(\kappa) = \beta(\kappa_0) + \sum\limits_{m\geqslant1}\frac{D_m}{m!}\Delta\kappa^m
		\label{eq:diffac_eqn}
	\end{align}
	where $D_m \equiv \left(d^m \beta/d \kappa^m\right)|_{\kappa_0}$ and $\Delta \kappa = \kappa -\kappa_0$.
	Performing a Fourier transformation to change the domain as $\kappa \rightarrow n$ by replacing $\Delta \kappa \equiv -i \partial_n$, where $n$ is defined as a continuous variable of an amplitude function $\Psi(n,\xi)=\psi_{n,\xi}\exp(-i\kappa_0n)$, we have an approximate standard nonlinear Schr\"{o}dinger equation(NLSE)\cite{tranDiffractiveResonantRadiation2013}
	\begin{align}
		\left[i \partial_{\xi}  + \sum\limits_{m\geqslant2} \frac{D_m}{m!}\left(-i \partial_n\right)^m + \abs{\psi(n,\xi)}^2\right]\psi(n,\xi) = 0	
		\label{eq:sNLSE}
	\end{align}
	Defining $n$ as a continuous variable can be justified by the fact that the involved solitons encompass several waveguides.
	By using the concept of co-moving frame $n\rightarrow n + D_1 \xi$ and introducing a phase evolution substitution $\psi(n,\xi) = \psi(n,\xi)\exp\left[i \beta(\kappa_0) n\right]$ we can eliminate the first and second term of the Taylor expansion resulting a soliton solution for Eq.\ref{eq:sNLSE} with $D_{m\geqslant3} = 0$ as,
	\begin{align}
		\psi_{sol} = \psi_0 \sech\left(\frac{n \psi_0}{\sqrt{\abs{D_2}}}\right) \exp\left(i k_{sol} \xi\right),
		\label{eq:sol_soln}
	\end{align}
	 here $k_{sol} \equiv \psi_0^2/2$ is the longitudinal wavenumber of the soliton.
	Note that a bright soliton exists only when condition $\abs{\kappa_0} < \pi/2$ or $2c\cos(\kappa_0)>0$ is satisfied.
	Fig.\ref{fig:sol_lattice}(a) describes such a  soliton forming and propagating in nonlinear uniform WA for an input beam $\psi_{sol}=\psi_0 \sech(n\psi_0/\sqrt{|D_2|})$. In Fig.\ref{fig:sol_lattice}(b) the spatial distribution of the  DS is illustrated in the background of periodic refractive index grid offered by the typical WA.
	Inserting the plane-wave solution $\exp\left[i\left(k_{lin}\xi + \Delta \kappa n\right)\right]$ in a linearized Eq.\ref{eq:sNLSE} we obtain the dispersion relation	$	k_{lin}(\Delta \kappa) = \beta(\kappa)- \beta(\kappa_0) - D_1 \Delta \kappa$. A soliton of the form in Eq.(\ref{eq:sol_soln}) emits a radiation in $\kappa$-space by transferring energy to the linear wave when the condition $k_{sol} = k_{lin}(\Delta \kappa)$ is satisfied. 
	This is the PM condition required for generating DifRR \cite{tranDiffractiveResonantRadiation2013}, which can be further expressed as,
	\begin{align}
		\left[\cos(\kappa) - \cos(\kappa_0) + \sin(\kappa_0)\Delta \kappa\right] = \widetilde{\psi_0}^2
		\label{eq:PM}
	\end{align}
	where $\widetilde{\psi_0} = \psi_0/2\sqrt{c}$.
	The solution to this relation gives the wavenumber of the generated DifRR ($\kappa_{RR} = \kappa_0 + \Delta\kappa$) as a function of the soliton wavenumber $\kappa_0$.
	In Fig. \ref{fig:sol_lattice}(c), formation of the DifRR in $n$-space is demonstrated with an arrow highlighting the evolution of DifRR. 
	The Signature of  DifRR (around -2.8) is prominent in $\kappa$-space as illustrated in Fig. \ref{fig:sol_lattice}(d) and the output spectrograme in Fig.\ref{fig:sol_lattice} (e).
It is to be noted that the generated DifRR is also subjected to same limits ($\kappa < \abs\pi$) within the first Brillouin zone and undergo a shift (formally knows as \textit{anomalous recoil}) in it's wavenumber by $\pm 2\pi$ when they form outside these limits.
	In Fig. \ref{fig:sol_lattice}(f) we plot the wavenumber ($\kappa_{RR}$) of DifRR as a function of initial soliton wavenumber $\kappa_0$.
	However, it is observed  that  $k_{RR}$ obtained from numerical simulation deviate from the result predicted by the PM condition when $k_0$ is relatively large.
	It can be noted that, the recoil of the soliton after emitting the radiation is significant for higher values of $k_0$ resulting a change in its wavenumber.
		We track this change of  soliton wavenumber as marked by the white dotted line in Fig. \ref{fig:sol_lattice} (f).
		We observe that, during propagation the soliton wavenumber undergoes a change from $\kappa_0$  to $\kappa'_0$. So we replace the average wavenumber  $\kappa_0 \rightarrow (\kappa_0'+\kappa_0)/2$ in the PM condition Eq. \ref{eq:PM} and find a better agreement with numerical results as shown by the solid black line in Fig. \ref{fig:sol_lattice} (f). 
		 Further we calculate  the fraction of energy ($\Delta E_{RR}=\int_{\kappa_i}^{\kappa_f} |\tilde{\psi}_{RR}(\kappa,\xi)|^2 d \kappa/\int_{-\pi}^{\pi} |\tilde{\psi}(\kappa,\xi)|^2 d\kappa$) that is accumulated in DifRR as a function of initial wavenumber ($\kappa_0$). The variation of $\Delta E_{RR}$ as a function of $\kappa_0$ is also depicted as a gray curve in \ref{fig:sol_lattice} (f).

		\subsection{Waveguide Array Design}
\begin{figure}[h]
		\begin{center}
			\includegraphics[width=\linewidth]{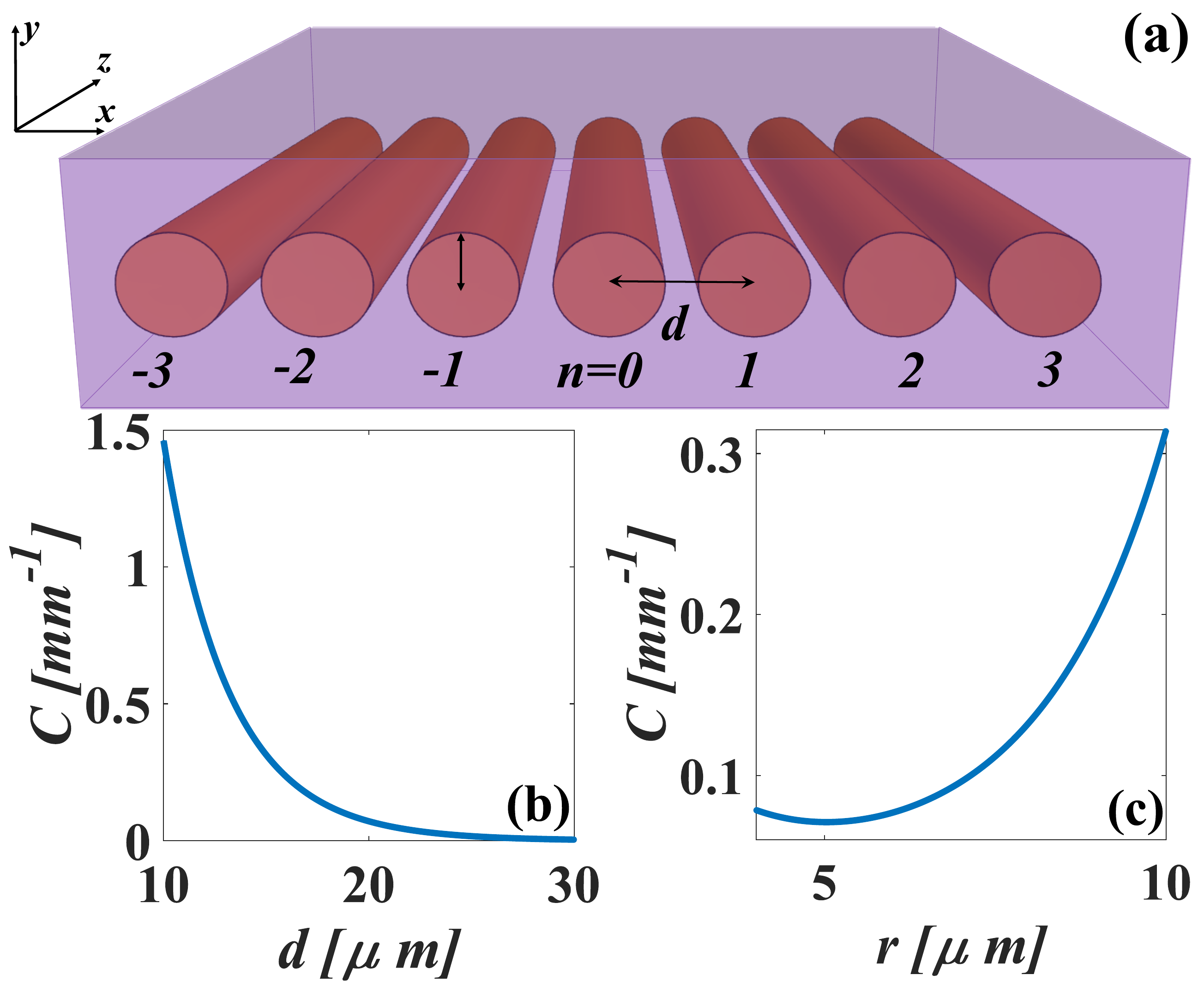}
		\end{center}
		\caption{(a) Schematic representation of the proposed WA with cylindrical cores of radius $r$, where adjacent waveguides are uniformly separated by $d$.
			(b) Variation of coupling coefficient as a function of separation between adjacent waveguides for $r = 5 \mu$m.  (c) Variation of coupling coefficient as a function of core radius for a fixed separation $d = 20 \mu$m.}
		\label{fig:wa_design}
	\end{figure}

	Before we proceed to a detailed analysis of collision dynamics, we define a physically realizable waveguide structure by utilizing the facility of fs laser based writing in transparent bulk media\cite{szameitHexagonalWaveguideArrays2006a,pavlovFemtosecondLaserWritten2017}.
	This enables us to prepare a WA composed of GeO$_2$ doped silica cores suspended in a silica cladding as illustrated in Fig.\ref{fig:wa_design} (a).
	The refractive indices of the core and cladding are $n_1 \approx 1.4477$ and $n_2 \approx 1.4446$, respectively, at the operating wavelength of $\lambda_0 = 1.55$ $\mu$m.
	We consider cylindrical cores with radius $r= 5$ $\mu$m, separated by a distance $d = 20$ $ \mu$m with a calculated nonlinear coefficient of $\gamma = 0.79$ W$^{-1}$km$^{-1}$ for the given geometry.
	For such a WA arrangement, the coupling coefficient between adjacent waveguides is given by\cite{tewariAnalysisTunableSinglemode1986}
	\begin{equation}
		C (d,r)= \frac{\lambda_0}{2 \pi n_1} \frac{U^2}{r^2 V^2}\frac{K_0(Wd/r)}{K_1^2(W)},
		\label{eq:coup_coeff}
	\end{equation}
	where $\lambda$ is the free space wavelength, and $K_{\nu}$ are the modified Bessel functions of the second kind of order $\nu$.
	$U$ and $V$ are the mode parameters that satisfy $U^2 + W^2 = V^2$, where the $V$ parameter is defined as $V = k_0r \sqrt{n_1^2 - n_2^2}$, and $U$ is approximated as $U \approxeq 2.405 e^{-(1-\Delta/2)/V}$, with $\Delta = 1- (n_2/n_1)^2$\cite{snyderCoupledModeTheoryOptical1972}.
	Here, $r$ is the core radius, $k_0(=2\pi/\lambda_0)$ is the free space wavevector and $d$ represents the separation between two adjacent waveguides. 
	The variation of the coupling coefficient as a function of separation $d$ for a fixed radius $r$ and vice versa is plotted in Fig.\ref{fig:wa_design}(b)(c). 
	In our numerical analysis we consider a homogeneous WA composed of waveguides with radius $r=$ 5 $\mu$m and separation $d=$ 20 $\mu$m. 
	For these parameters we obtain the value of the coupling coefficient to be $\approxeq$ 0.07 mm$^{-1}$.

	\subsection{Soliton Collision in DNLS Systems}
	Solitons by nature maintain their shape and location during propagation.
	This property is also known to extend to collisions between two solitons in continuous domain where they interact elastically and appear to pass through each other.
	After such a collision the solitons  undergo an instant translation in space and/or time accompanied by a phase shift.
	This property was originally observed in a study of Korteweg-de-Vries (KdV) equation and later in systems governed by NLSE\cite{aosseyPropertiesSolitonsolitonCollisions1992}.
	While two solitons in close vicinity interact with an attractive or repulsive potential based on their relative initial phase, the strength of this interaction potential is however very weak to act over a large separation in $n$-space.
	Hence, for two solitons separated by a large enough distance are to be provided a transverse motion for them to interact.
	In this case, the soliton is initiated by a \enquote{\textit{push}} in the form of a phase gradient $k$ across the solitons at their inputs\cite{papacharalampousSolitonCollisionsDiscrete2003}.
	Two solitons with appropriate opposite signs of phase gradient $k$ can be made to collide with their transverse velocities proportional to $k$.
	Such a pair of DSs at the input is given by,
	\begin{align}
		\begin{split}
			\psi(n,0) = A_1 \sech\left[\frac{A_1 (n- n_{1})}{\sqrt{2c\cos{k_1}}}\right]\exp\left[ik_1(n-n_1)\right] \\ 
			+ A_2 \sech\left[\frac{A_2 (n- n_{2})}{\sqrt{2c\cos{k_2}}}\right]\exp\left[ik_2(n-n_2)\right],
		\end{split}
		\label{eq:two_sol}
	\end{align}
	where $n_{j=1,2}$ are the location of the DS peaks, $A_{j=1,2}$ are the amplitudes and $k_{j=1,2}$ are the respective wavenumbers.
	$k_{j=1,2}$ are real and their signs determines whether the solitons will travel in opposite directions or towards each other while they propagate.
	One valid combination for soliton collision is $k_1,n_2 <0$ and  $k_2,n_1>0$.
	It is to be noted that the reversal of their signs is also a valid combination, however the results obtained merely mirror the phenomenon observed in the initial case.
	The DSs after a collision either emerge with their properties intact or fuse together to form a \textit{breather}\cite{sakaiSolitoncollisionInterferometerQuantum1990,aosseyPropertiesSolitonsolitonCollisions1992}.
	The results of the collision are determined and controlled by the soliton amplitude $A_j$ and wavevector $k_j$.
	In subsequent sections we will analyse the deeper aspects of the collision dynamics and other related phenomena  along with the results of numerical analysis.

	\section{Numerical and analytical results}
	In this section we numerically investigate the results of interaction between two DSs defined by Eq.\ref{eq:two_sol} in a uniform homogeneous WA described in Fig.\ref{fig:wa_design} (a).
	We consider a WA with the range of it's $n$ index as $-350\leqslant n \leqslant 350$ which defines $N=350$ and a total of $2N + 1 = 701$ waveguides in the array. We also develop an analytical treatment based on the variational technique to grasp the collision mechanism between DS pair.

	\subsection{Collision of two identical DSs}
\begin{figure}[h]
		\begin{center}
			\includegraphics[width=1.0\linewidth]{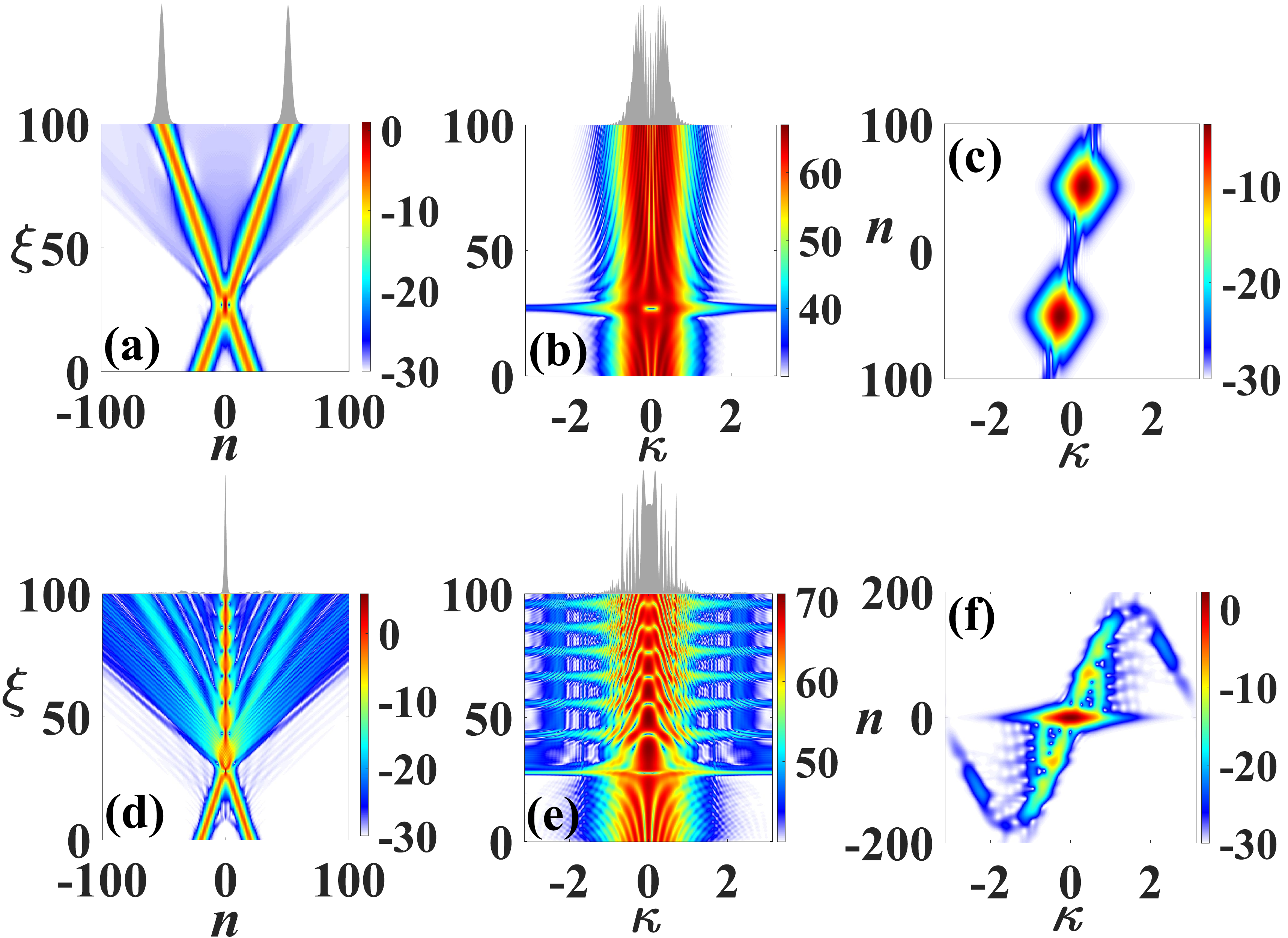}
		\end{center}
		\caption{Evolution of two identical DS pair launched in a WA obliquely with $k_0 = 0.3$ towards each other. Evolution in (a) $n$ and (b) $\kappa$ domains for amplitude $A_0 = 0.5$ result in a crossover state followed by an elastic like collision.  (c) Spectrogram at output ($\xi = 100$) showing two distinct DSs in ($n$-$k$ space). 
(d) Two DS of amplitude $A_0 = 0.8$ launched with the same value of $k_0$ merge together to form a \textit{breather} with small side-lobes. (e) Evolution of the breather in $\kappa$-space. (f) Spectrogram at output showing the single merge state where the side-lobes are distributed in $n$ and $\kappa$ domain in the form of multiple weak radiation. In all the heat maps  $|\psi_n|^2$ is plotted in log scale along $z$-axis .}
		\label{fig:2_sol_col}
	\end{figure}

\begin{figure}[h]
		\begin{center}
			\includegraphics[width=1.0\linewidth]{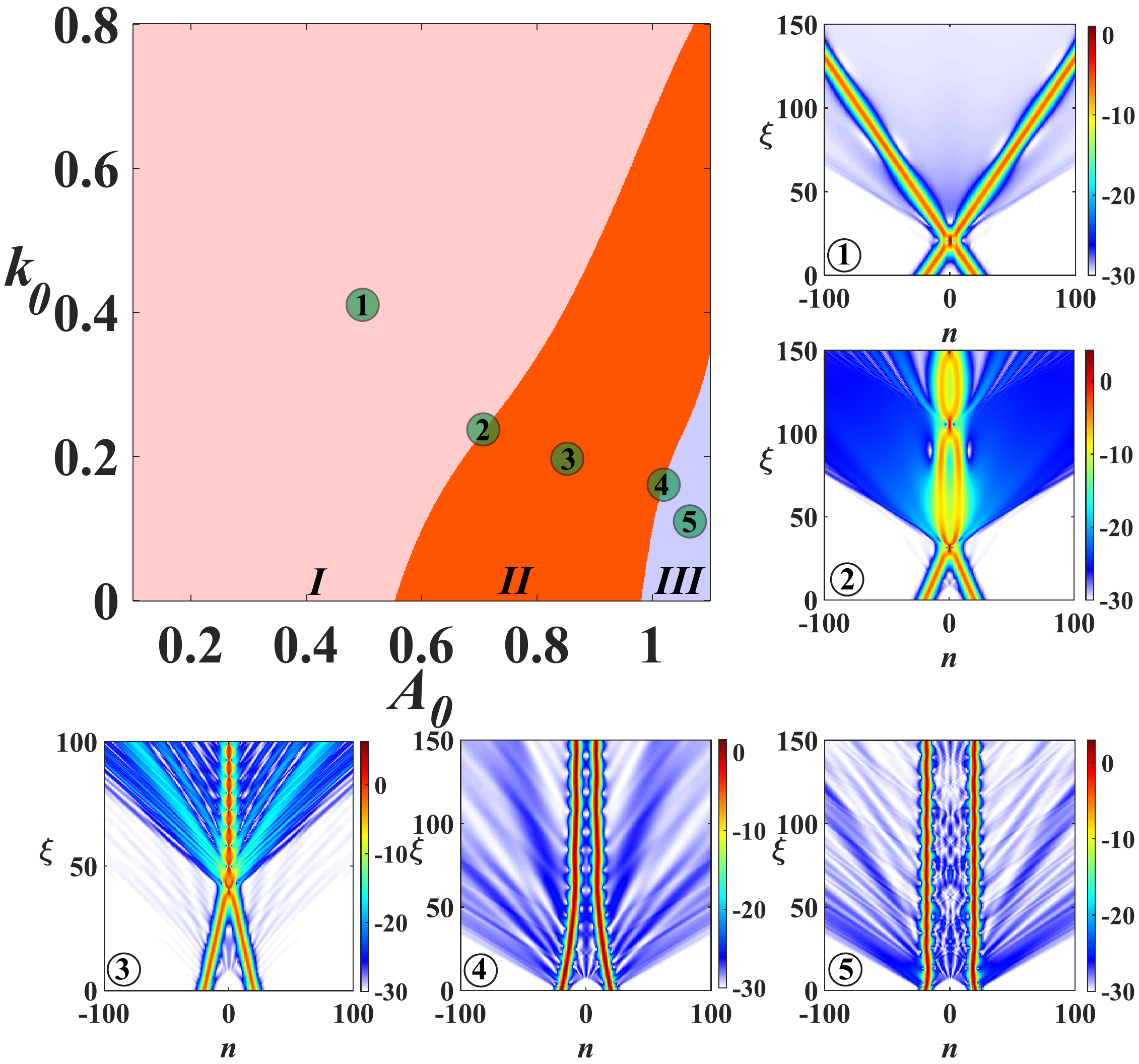}
			\caption{A phase map of the interaction between two identical DSs as function of amplitude ($A_0$) and wavenumber ($\kappa_0$), accompanied by sub-figures describing the evolution of DSs at different regime.
					Region (\textit{I}) and (\textit{II}) in the phase map correspond to elastic collision and breather formation respectively, with \circled{2} being the boundary between these two domains.
					Any soliton with values of $A_0$ and $k_0$ in region (\textit{III}) are subjected to a \textit{stop band} and the solitons are bound to their launching sites as seen in the case of \circled{5}.  In all the heat maps, along $z$-axis $|\psi_n|^2$ is plotted in log scale for  better resolution.}	
			\label{fig:breather_phase}
		\end{center}
	\end{figure}

	A pair of identical DSs equidistant from central waveguide ($n=0$) are defined by taking,  $n_0 = n_1 = -n_2$,  $k_0 = k_2 = -k_1$ and $A_{j=1,2} = A_0$, in Eq.\ref{eq:two_sol}.
	The evolution and collision dynamics of such a pair for DSs with initial wavenumber $k_0 = 0.3$, starting at $n_0 = 20$ for two different amplitudes $A_0 = 0.5$ and $0.8$,  are illustrated in Fig.\ref{fig:2_sol_col}.
	In Fig.\ref{fig:2_sol_col}(a),(b) we observed the evolution of the soliton pair in the $n$ and $\kappa$ domain, respectively for $A_0 = 0.5$, where they interact and emerge while maintaining their shape and properties under an elastic like collision.
	Numerically, we find the energy carried by each DS, $E = 2A_0\sqrt{2c \cos(k_0)} = 1.514$ remains conserved throughout the propagation.
	The spectrogram plot in Fig.\ref{fig:2_sol_col} (c) at output reflects how two DSs remain intact followed by an elastic like collision. To visualize the complete dynamics see Supplemental Material \cite{movie1}
	The DS pair are fused together to form a discrete \textit{breather} when we increase the amplitude to $A_0 = 0.8$ (keeping $k_0=0.3$, same as before).  The formation of this fused state  is shown in Fig.\ref{fig:2_sol_col} (d) and (e). The spectrogram in ($n$-$k$) space, as shown in Fig.\ref{fig:2_sol_col} (f), depicts the formation of a single state with side-lodes exhibiting multiple weak radiations. To visualize the complete picture see Supplemental Material \cite{movie2}. The amplitude $A_0$ and initial wavenumber $k_0$ determines whether DS pair will collide elastically or form a \textit{breather}. The DSs even do not collide and form a bound-state in the limit $A_0>1, k_0<0.1$. 
	By compiling the results for over a range of $A_0$ and $k_0$ we develop a phase diagram in Fig.\ref{fig:breather_phase} showing the formation of three distinct states namely \textit{I}, \textit{II} and \textit{III} corresponding to elastic collision, \textit{breather} and \textit{bound-state} formation, respectively. 
	
\begin{figure}[h]
	\begin{center}
		\includegraphics[width=1.0\linewidth]{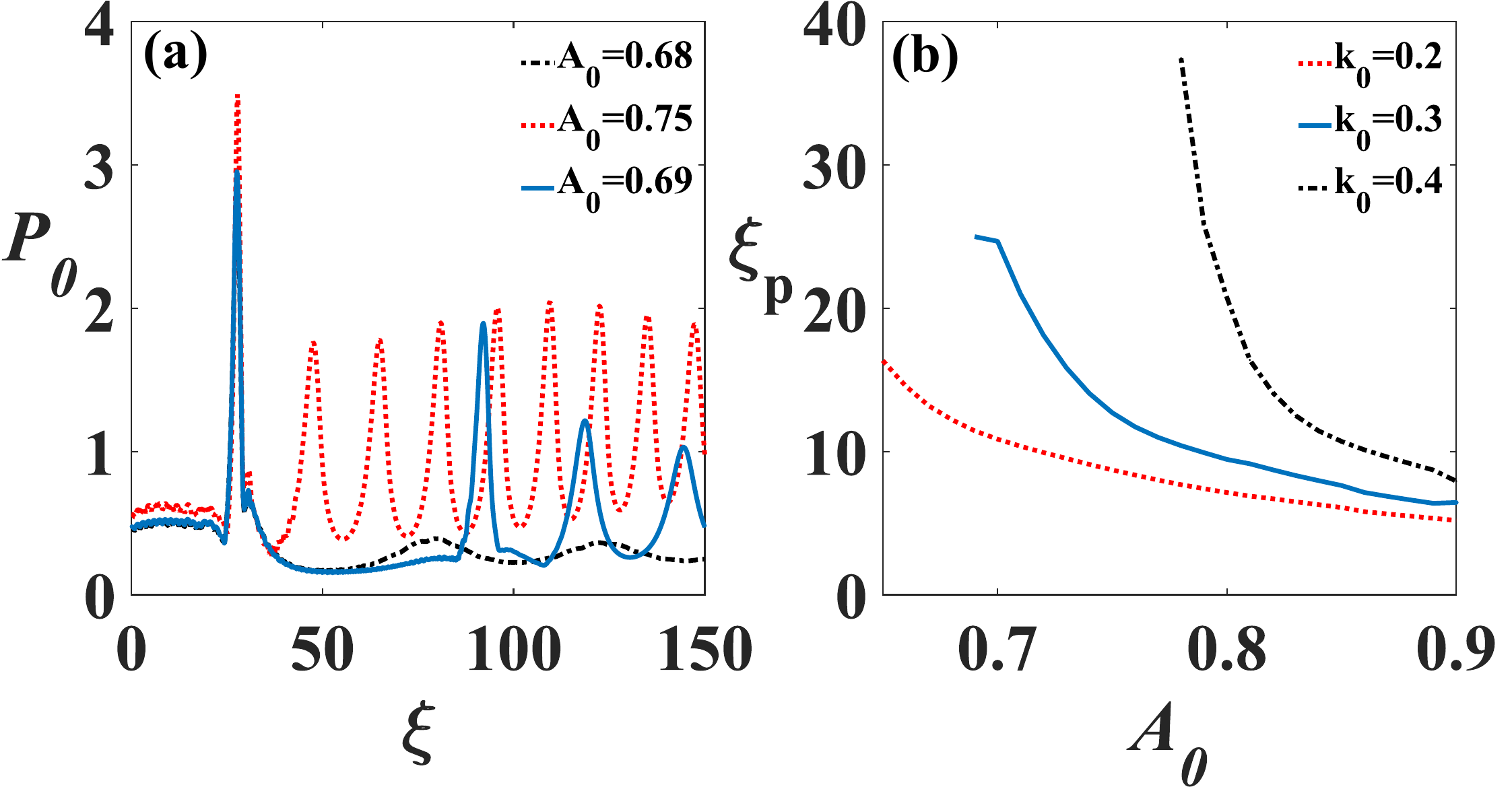}
		\caption{(a) Peak power ($P_0$) is plotted as a function of $\xi$ for different input amplitude ($A_0$) and  $k_0 = 0.3$.  The largest value corresponding to location of soliton interaction. The periodicity ($\xi_p$) differs for crossover state (blacked dashed), breather state (red dotted) and boundary regions (solid blue). From the comparison it is evident that near boundary the periodicity of the breather is larger.
				(b) \textit{Breather} period $\xi_p$ as a function of $A_0$ for three different initial wavenumbers $k_0$ equals $0.2$ (red dotted), $0.3$ (solid blue), and $0.4$ (black dashed). $\xi_p \approx \pi$ for all the \textit{breathers} near \textit{stop-band}. }
		\label{fig:peak_osc}
	\end{center}
	\end{figure}

	The  evolution of soliton pairs at phase locations, \circled{1} ($k_0 = 0.4$,$A_0=0.5$) for elastic collision and \circled{3} ($k_0 = 0.2$,$A_0 = 0.85$) for \textit{breather} formation are depicted in the sub-plots of Fig. \ref{fig:breather_phase}. We also illustrate the collision dynamics at the phase boundary  \circled{2} for  $k_0= 0.253$, $A_0=0.65$ showing \textit{breather} formation with large period. 
	The region \textit{III}  corresponds to a \textit{stop band} where solitons do not interact \cite{papacharalampousSolitonCollisionsDiscrete2003}.
	It is observed that, for a normalized coupling coefficient $c = 1.2$, this \textit{stop band} begins to appear for low values of $k_0$ and high $A_0$. 
	In the  \textit{stop band} region\circled{5} the soliton pair forms a bound state as illustrated in the sub-plot of Fig. \ref{fig:breather_phase}. 
	Note that, for an exact soliton solution input,  $\psi_{sol}=\psi_0\sech(n\psi_0/\sqrt{|D_2|})$ we do not observe any symmetry-breaking behavior as observed in earlier work \cite{papacharalampousSolitonCollisionsDiscrete2003}.
	We also demonstrate the soliton dynamics at phase-boundary \circled{4} for the parameters  $k_0=0.15$, $A_0=1.035$. 
	Any solitons with the same $k_0$ ($A_0$) and higher (lower) $A_0$ ($k_0$) are subjected to this \textit{stop band}.
	 The formation of breather is characterized by the periodic evolution of its peak power over propagation distance. Separation between the consecutive peaks  is defined as period ($\xi_p$) of the \textit{breather} that depends on the soliton parameter $k_0$ and $A_0$. In Fig. \ref{fig:peak_osc}(a) we plot the variation of the peak power $P_0$ as a function of $\xi$ for different $A_0$ to visualize the relative periodicity of the \textit{breather}.  It is evident that the periodicity $\xi_p$ is sensitive to $A_0$. To grasp the whole picture we plot $\xi_p$ as a function of $A_0$ in  Fig. \ref{fig:peak_osc}(b) for three different $k_0$. It is observed that, for any given $k_0$ the  \textit{breather} starts with a relatively high period and gradually decreases until the \textit{stop-band} appears. It is interesting to note that, for the given coupling coefficient ($c=1.2$), the periodicity  $\xi_p$ of all the \textit{breathers} formed near the \textit{stop-band} is almost equal, $\xi_p\rightarrow \pi$.
	\noindent
	
		\noindent

	To shed more light on the collision dynamics,  we further study the evolution of relative phase of the DSs (and \textit{breather}) along propagation distance.
	Here we consider the initial phase detuning of the soliton pair to be $0$, $i.e.$ DSs are in-phase. 
	In time domain, the average soliton phase at a point in the propagation axis ($\xi$) is defined as  \cite{blowSolitonPhase1992a},
	\begin{align}
		\phi(\xi) = \tan^{-1}\left[\frac{\int |\psi(t,\xi)|^2 \Im[\psi(t,\xi)] dt}{\int |\psi(t,\xi)|^2 \Re[\psi(t,\xi)] dt}\right],
		\label{eq:av_phase}
	\end{align}
	with the contribution to the average phase being limited to near pulse electric field by a pulse intensity weight $\abs{\psi(\xi,t)}^2$\cite{blowSolitonPhase1992a}.
	We consider the fact that the NLSE in spatial domain is analogous to  its temporal counterpart and rewrite this equation in terms of $n$ ($t\rightarrow n$).
	Discretizing  the Eq. (\ref{eq:av_phase}) by the approximation $dn\rightarrow 1$, we obtain.
	\begin{equation}
		\phi(\xi) = \tan^{-1}\left[\frac{\sum\limits_{n} |\psi_n(\xi)|^2 \Im(\psi_n(\xi))}{\sum\limits_{n} |\psi_n(\xi)|^2 \Re(\psi_n(\xi))}\right].
		\label{eq:av_phase_n}
	\end{equation}
Note, the average \enquote{lattice} phase for a pure DS (with $k_0=0$) oscillates between $\pi/2$ to $-\pi/2$  as it progresses along $\xi$.
	This is attributed to the range $\left(-\pi/2,\pi/2\right)$ of the $\tan^{-1}$ function.
	This repetition is periodic in a case of a standard DS, and on unwrapping the value of $\phi(\xi)$ i.e adding $\pi$ to the value of $\phi(\xi)$ whenever it emerges from the other side of the domain, we obtain $\phi(\xi)$ as a linear function of $\xi$\cite{blowSolitonPhase1992a}.
	We similarly calculate the average phase of the system for both the cases of elastic collision and \textit{breather} formation ($\phi_k(\xi)$), then take the difference between it and that of the pure DS ($\phi_0(\xi)$) which is considered to be a reference phase. 
	The evolution of the relative phase $\phi_k(\xi)-\phi_0(\xi)=\Delta \phi(\xi)$ helps us to achieve a better resolution if any phase change occurs during the interaction between the DS pair.
	\begin{figure}[h]
		\begin{center}
			\includegraphics[width=1.0\linewidth]{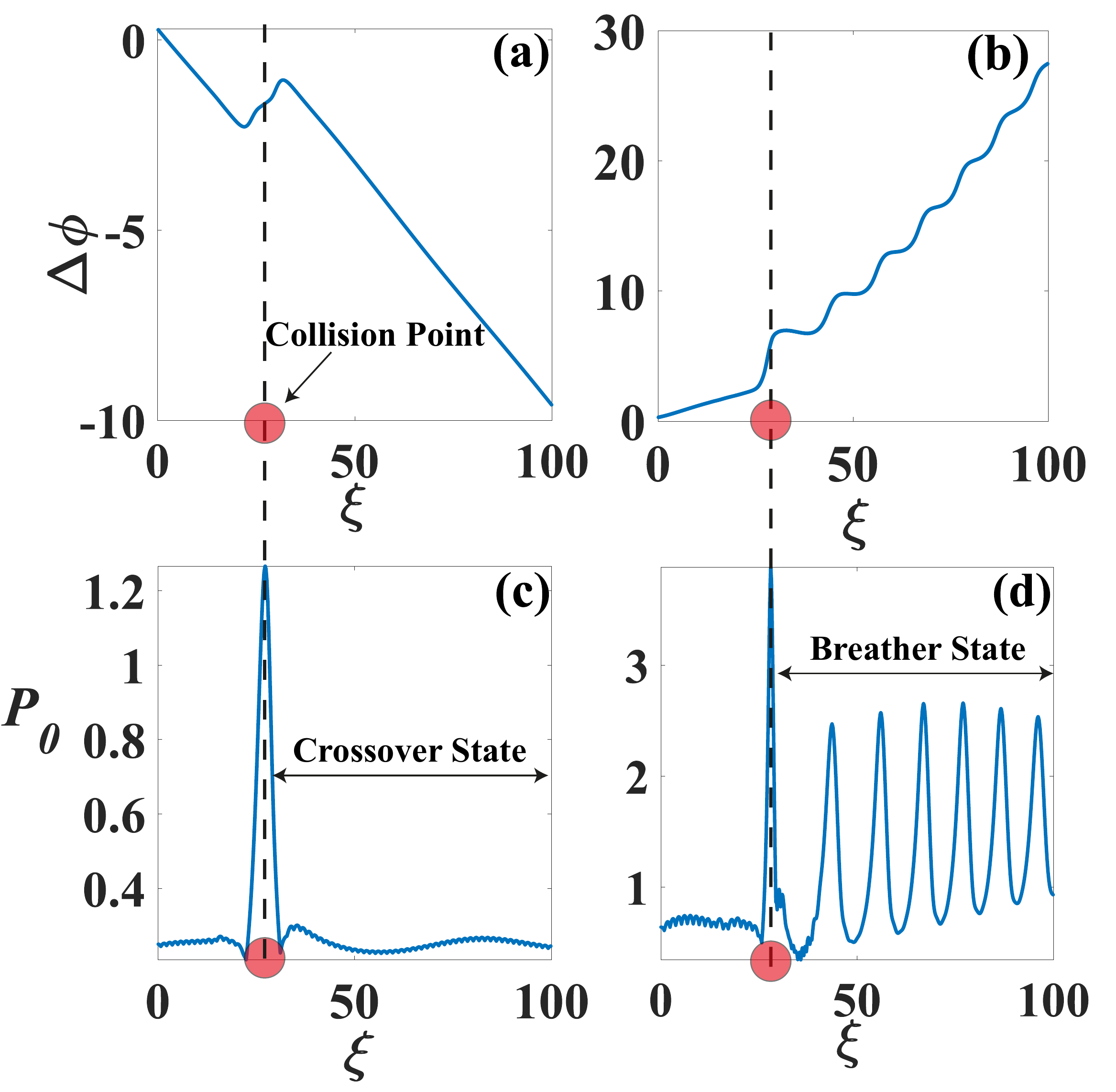}
		\end{center}
		\caption{Evolution of (a),(b) relative phase and (c),(d) peak power for elastic collision [(a) and (c)] and breather formation [(b) and (d)]. Point of collision ($\xi_{col}$) is accompanied by a spike in (c) peak power and (a) abrupt change in phase evolution.
		 For a \textit{breather} formation, peak power oscillates periodically (d) reciprocating (b)the step-wise increase in phase. Point of collision at $\xi$ is represented by a dotted line and red circle in each case.}
		\label{fig:phase_evo_b_nb}
	\end{figure}
	We simultaneously calculate the evolution of the peak power in the system which is defined as $P_0 (= |\psi_{max}|^2)$, where $\psi_{max}$ is the maxima of the total field in the WA.
	Fig.\ref{fig:phase_evo_b_nb}(a), (c) and (b),(d) show the variation of $\Delta \phi$  and $P_0$ (in same frame) for elastic collision and \textit{breather} formation, respectively. It is evident that $\Delta \phi$  shifts abruptly at collision point. For fused state,  periodic \enquote{lumps} are observed in relative phase evolution.

\noindent

\subsection{Variational Analysis}

\begin{figure}[h]
	\includegraphics[width=1.0\linewidth]{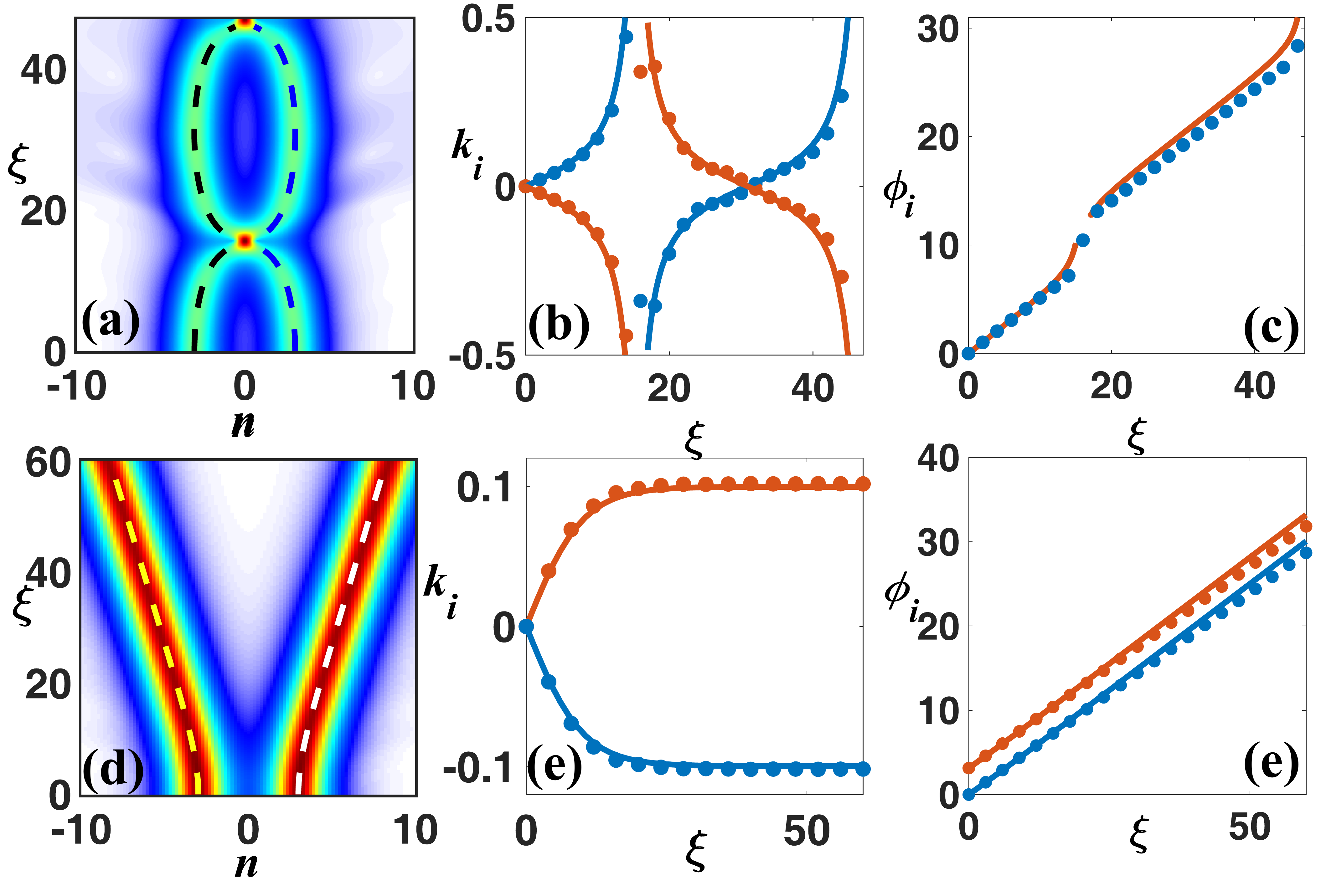}
	\caption{(a),(d) Collision dynamics of in-phase ($\Delta \Phi =0$) and out-of-phase ($\Delta \Phi =\pi$) soliton pair (b),(e) wavenumber and (c),(f) phase variation, respectively. The dashed lines in (a) and (d) represent soliton trajectory which we obtain from variational result. The solid lines in other figures are variational results which corroborate well with numerical simulation (solid dots). }
	
	\label{vafig}
\end{figure}

The assumption of continuous transverse variable ($n$) allow us to exploit the variational analysis for soliton collision problem where  we write Eq. \ref{eq:sNLSE} in the form of two perturbed coupled NLSEs,
\begin{equation}
	i \partial_{\xi} \psi_{\ell} + \frac{1}{2}\partial_{n}^2 \psi_{\ell} + |\psi_{\ell}|^2 \psi_{\ell}  = i\epsilon_{\ell}
	\label{eq:2_sol_NLSE} .
\end{equation}
Here $\psi_{\ell=1,2}$ represents two soliton fields 1 and 2, coupled by the perturbation $\epsilon_{\ell} = i[2 |\psi_{\ell}|^2 \psi_ {3-\ell} + \psi_{\ell}^2 \psi_{3-\ell}^*]$ which one can obtain by replacing $\psi \rightarrow \psi_1+\psi_2$ in Eq. \ref{eq:sNLSE} . By introducing the Lagrangian density $\mathcal{L}_D = i/2 (\psi_{\ell} \partial_{\xi} \psi_{\ell}^* - \psi_{\ell}^* \partial_{\xi}\psi_{\ell}) + 1/2|\partial_{n}\psi_{\ell}|^2 - 1/2|\psi_{\ell}|^4 + i(\epsilon_{\ell} \psi_{\ell}^* - \epsilon_{\ell}^* \psi_{\ell})$, appropriate for Eq. \ref{eq:2_sol_NLSE} and selecting a suitable  \textit{ans\"{a}tz} function $ \psi_{\ell} = \mathcal{A}_{\ell} \sech[\mathcal{A}_{\ell} (n - n_{\ell})]\exp[i \phi_{\ell} - i k_{\ell}(n - n_{\ell})]$,  we can reduced the Lagrangian $L = \int_{-\infty}^{\infty} \mathcal{L}_D d n$ :
\begin{equation}
	\begin{aligned}
		L_{\ell} = 2 \mathcal{A}_{\ell}  \frac{\partial \phi_{\ell}}{\partial_{\xi}} + 2\mathcal{A}_{\ell} \delta_{\ell} \frac{\partial n_{\ell}}{\partial_{\xi}} - \frac{\mathcal{A}_{\ell}}{3} + \mathcal{A}_{\ell} \delta_{\ell}^2 \\ - 2 \Im \int_{-\infty}^{\infty} ( \epsilon_{\ell} \psi_{\ell}^*) d n.
	\end{aligned}
\end{equation}
Using the Euler-Lagrange equation, we obtain the following set of eight coupled ODEs that describe the evolution of different parameters (like amplitude, position, wavenumber, phase) for both the solitons ($\ell=1,2$) ,
\begin{subequations}
	\begin{eqnarray}
		\frac{d \mathcal{A}_{\ell}}{d\xi} = (-1)^{\ell} \mathcal{A}_{\ell}^2 \mathcal{A}_{3-\ell} \sin(\Phi) \mathcal{F}_{\ell}\\
		\frac{d n_{\ell}}{d \xi} = - k_{\ell}  -(-1)^{\ell}  n_{\ell} \mathcal{A}_{\ell} \mathcal{A}_{3-\ell} \sin (\Phi) \mathcal{F}_{\ell}\\
 		\begin{aligned}
			\frac{d k_{\ell}}{d \xi} = - \mathcal{A}_{\ell} \mathcal{A}_{3-\ell}^2 \cos (\Phi) \mathcal{F}'_{n_{\ell}} 
		\end{aligned}\\
		\begin{aligned}
			\frac{d \phi_{\ell}}{d \xi} = \frac{1}{2}(\mathcal{A}_{\ell}^2 + k_{\ell}^2) -  (-1)^{\ell}  \mathcal{A}_{\ell} \mathcal{A}_{3-\ell} [k_{\ell} \sin (\Phi)\\ - 2 \cos (\Phi)] \mathcal{F}_{\ell},
		\end{aligned}
	\end{eqnarray}
	\label{VA}
\end{subequations}
\noindent where $\mathcal{F}_{\ell} = \csch^3 (\alpha_{\ell}) [\sinh (2\alpha_{\ell}) - 2\alpha_{\ell}]$, with $\alpha_{\ell} = \mathcal{A}_{3-\ell} (n_{\ell} - n_{3-\ell})$, $\Phi = (\phi_2 - \phi_1 )$ and $\mathcal{F}'_{n_{\ell}} =\frac{\partial\mathcal{F}_{\ell}}{\partial n_{\ell}}$.
This set of ODE \eqref{VA} provides valuable physical insights of the soliton collision problem. For example, the relative phase $\Phi$ appears in all the equation, $i.e$ initial phase detuning should significantly influence the collision dynamics. We dedicate the following section where we investigate the role of initial phase detuning in the context of soliton collision. In Fig. \ref{vafig} we compare our variational results with full numerical simulation. Variational results nicely predict the evolution of soliton parameters under collision. As expected, abrupt change in wavenumber is noticed (see Fig. \ref{vafig} (b)) at the point of interaction due to elastic collision. The sudden chance in the phase is also noticed (see Fig. \ref{vafig} (c)) which is consistent with our earlier result shown in Fig. \ref{fig:phase_evo_b_nb} (a).

\begin{figure}[h]
	\includegraphics[width=0.9\linewidth]{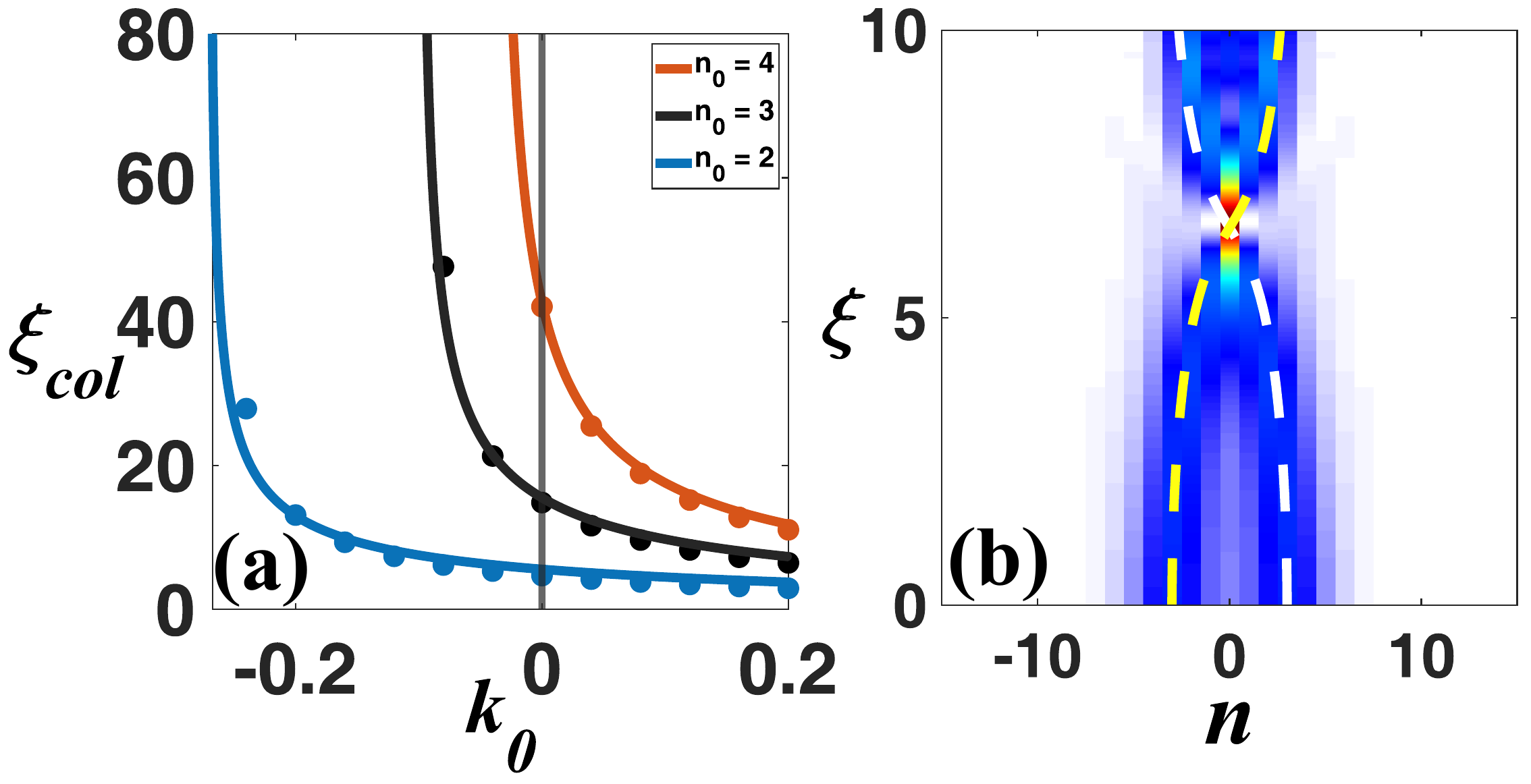}
	\caption{(a) Variation of collision point ($\xi_{col}$) as a function of initial wavenumber ($k_0$) of soliton pair when they are launched at waveguide number $n_0$, either side of the central waveguide. The numerical data are represented by solid dots where solid lines stand for variational prediction. (b) The variational analysis (dotted line) nicely predicts the trajectory of DS pair which we obtain by solving DNLSE Eq. \ref{eq:nDNLSE}.  }
	
	\label{vafig1}
\end{figure}

To establish the validity of the variational analysis for wider range of parameters, we numerically calculate the location of first collision point $\xi_{col}$ (for a given $n_0$) as a function of initial wavenumber $k_0$ and try to match it with variational prediction. As shown in Fig. \ref{vafig1}(a)	the variational results that we obtain by solving Eq. \ref{VA} corroborate well with numerical data. As expected the collision point decreases with increasing $k_0$. 
		
Finally in Fig.\ref{vafig1}(b) we superimpose the variational result with the collision dynamics of DS that we obtain by solving the pure DNLSE Eq. \ref{eq:nDNLSE}. Here we include the coupling coefficient in the variational analysis and obtain a satisfactory match.

		\subsection{Role of initial phase detuning in soliton collision:}

\begin{figure}[h]
	\begin{center}
		\includegraphics[width=1.0\linewidth]{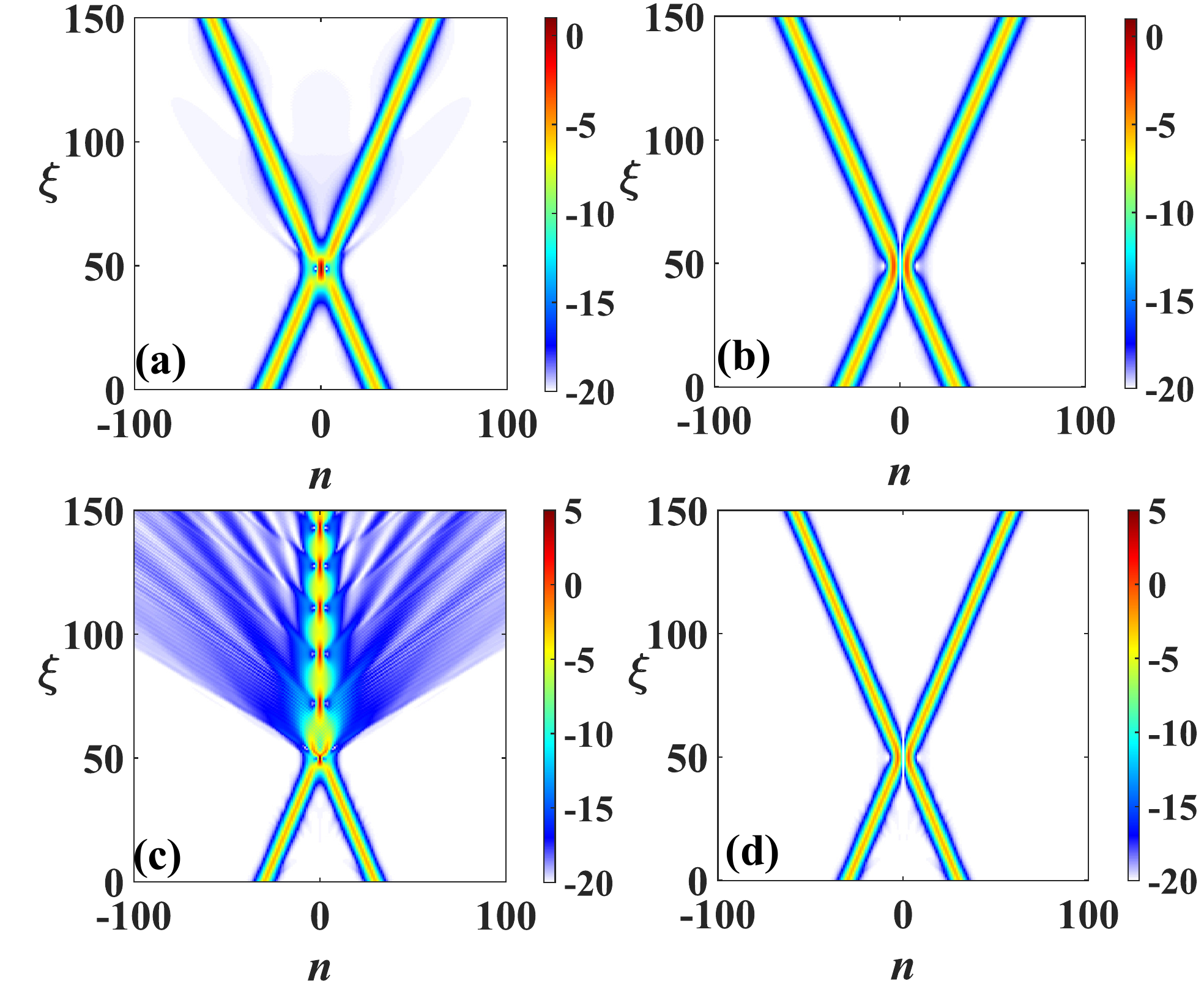}
		\caption{Collision of two DSs with initial phase difference $\Delta \Phi = 0$ (a) and (c), $\Delta \Phi = \pi$ (b) and (d). DSs in figures (a) and (b) are launched with parameters $A_0 = 0.5$, $k_0 = 0.25$ lying in the domain of elastic collision, with (c) and (d) $A_0 = 0.7$,$k_0 = 0.25$ corresponding to breather formation.
				DSs with phase difference of $\pi$ exhibit the absence of \textit{breather} formation. In all the heat maps  $|\psi_n|^2$ is plotted in log scale along $z$-axis for better clarity.}
		\label{fig:pd_col}
	\end{center}
\end{figure}

\begin{figure}[h]
		\begin{center}
			\includegraphics[width=0.9\linewidth]{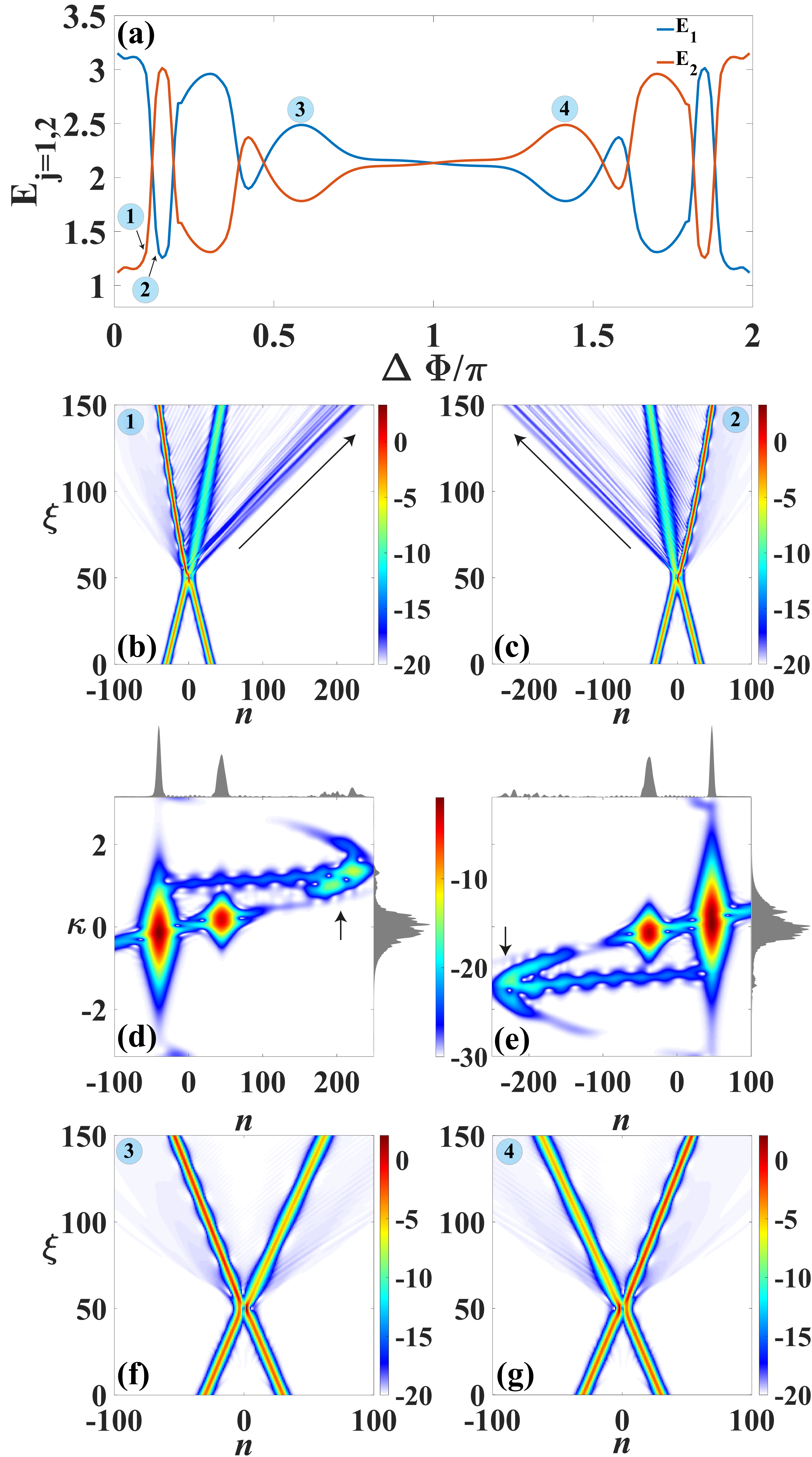}
			\caption{(a) Energy ($E_{j=1,2}$) exchange among the two DS $\psi_{j=1,2}$ as a function of initial phase detuning where  $A_0 = 0.7$, $k_0 = 0.25$. Plot (b),(c) represent the collision dynamics for the phase detuning ($\Delta \Phi / \pi$), 0.10 and 0.14 as marked by \circled{1} and \circled{2}, respectively. The arrows show the formation of weak DifRR in $n$-space. Plot (d),(e) illustrate the  spectrogram of (b) and (c), respectively where the weak collision mediated radiation is evident as marked by arrows. Plot (f),(g) represent the collision dynamics for the phase detuning ($\Delta \Phi / \pi$), 0.59 and 1.41 as marked by \circled{3} and \circled{4}.}
			\label{fig 8}
		\end{center}
	\end{figure}

It is well established that, the initial phase plays a dominant role in two soliton interaction process when they are sufficiently close. Attractive and repulsive interactions are observed when the relative phase difference between two input solitons are $0$ and $\pi$, respectively. In order to understand the role of initial phase in the collision dynamics of DS pair, we allow the following field to propagate in the WA.

	\begin{align}
		\begin{split}
			\psi(n,0) = 
			A_0 \sech \left[\frac{A_0(n+n_0)}{\sqrt{2 c \cos k_0}}\right] \exp{\left[-ik_0(n+n_0)\right]} \\
			+ A_0 \sech \left[\frac{A_0(n-n_0)}{\sqrt{2 c \cos k_0}}\right] \exp{\left[ik_0(n-n_0)\right]\exp\left[i\Delta\Phi\right].}
		\end{split}
		\label{eq:pd_sol}
	\end{align}
	 Here, the total field is constructed by two solitons as, $\psi=\psi_1+\psi_2 $, where the initial phase difference between $\psi_1$ and $\psi_2$ is given by $\Delta\Phi$ as shown in Eq. (\ref{eq:pd_sol}).
	In our simulation we consider two set of ($A_0,k_0$) correspond to the region \textit{I} and \textit{II} in the phase diagrame. For $\Delta \phi=0$ we have the usual elastic collision and \textit{breather} formation as depicted in Fig.\ref{fig:pd_col} (a)  and (c), respectively. For the same set of parameter under out of phase condition, $i.e.$ $\Delta\Phi = \pi$ the DSs experience a repulsive interaction  similar to the temporal solitons with a phase difference of $\pi$.
		This repulsive behavior is even prominent for solitons with $A_0$ and $k_0$ in the regime \textit{II} suggesting that $\Delta\Phi = \pi$ prohibits the formation of \textit{breather}. It is also observed that the initial phase difference $\Delta \Phi$ doesn't play any role for the soliton pair in the \textit{stop band} region. It is worthy to note that, the \textit{breather} formation of two in-phase DSs ($\Delta \Phi=0$) is greatly influenced when $\Delta \Phi\neq0$. We scan the collision dynamics of two identical DS near phase boundary by continuously detuning the relative phase between them in the range $\Delta \Phi \rightarrow 0 - 2\pi$. We observe that the interaction between the two DSs $\psi_1$ and $\psi_2$ leads to a periodic energy exchange  against the detuned phase. In Fig. \ref{fig 8} (a) we demonstrate exchange of energy ($E=\sum |\psi_n|^2$) against initial phase detuning ($\Delta \Phi/\pi$) when two identical DSs are launched with equal amplitude $A_0= 0.7$ and wavenumber $|k_0|=0.25$.

	The collision dynamics of DS pair for four different detuned  phase are illustrated in  Fig. \ref{fig 8} (b),(c),(f),(g) where the energy exchange is evident and consistent with the experimental results given in \cite{stegemanOpticalSpatialSolitons1999a}.
	In  Fig. \ref{fig 8} (d), (e) we demonstrate the XFROG spectrogram corresponding to the plots Fig. \ref{fig 8} (b),(c) and observe a signature of weak DifRR in $\kappa$-space. This weak DifRR is originated due to the collision of DS pair and observed when the energy exchange between the DSs is relatively large (see, \circled{1} and \circled{2}).

	\subsection{Collision of two non-identical DS}
	Collision of identical DS is a special case, where the solitons are exactly identical and have the same and opposite transverse velocities.
	The natural extension to this case is a generalized study of two in phase DSs having unequal amplitudes and different initial wavenumbers as described in Eq.\ref{eq:two_sol} .
	We split this study into two parts, (a) solitons with equal amplitudes ($A_1 = A_2$) and nonidentical wavenumbers ($\abs{k_1} \neq \abs{k_2}$), and (b) both parameters of the solitons are unequal ($A_1 \neq A_2$ \& $\abs{k_1} \neq \abs{k_2}$).
	At a glance, it is evident from the soliton solution that solitons with non-identical wavenumbers  will have different widths proportional to $\sqrt{\cos (k_{j=1,2})}$.
	\begin{figure}[h]
		\begin{center}
			\includegraphics[width=1\linewidth]{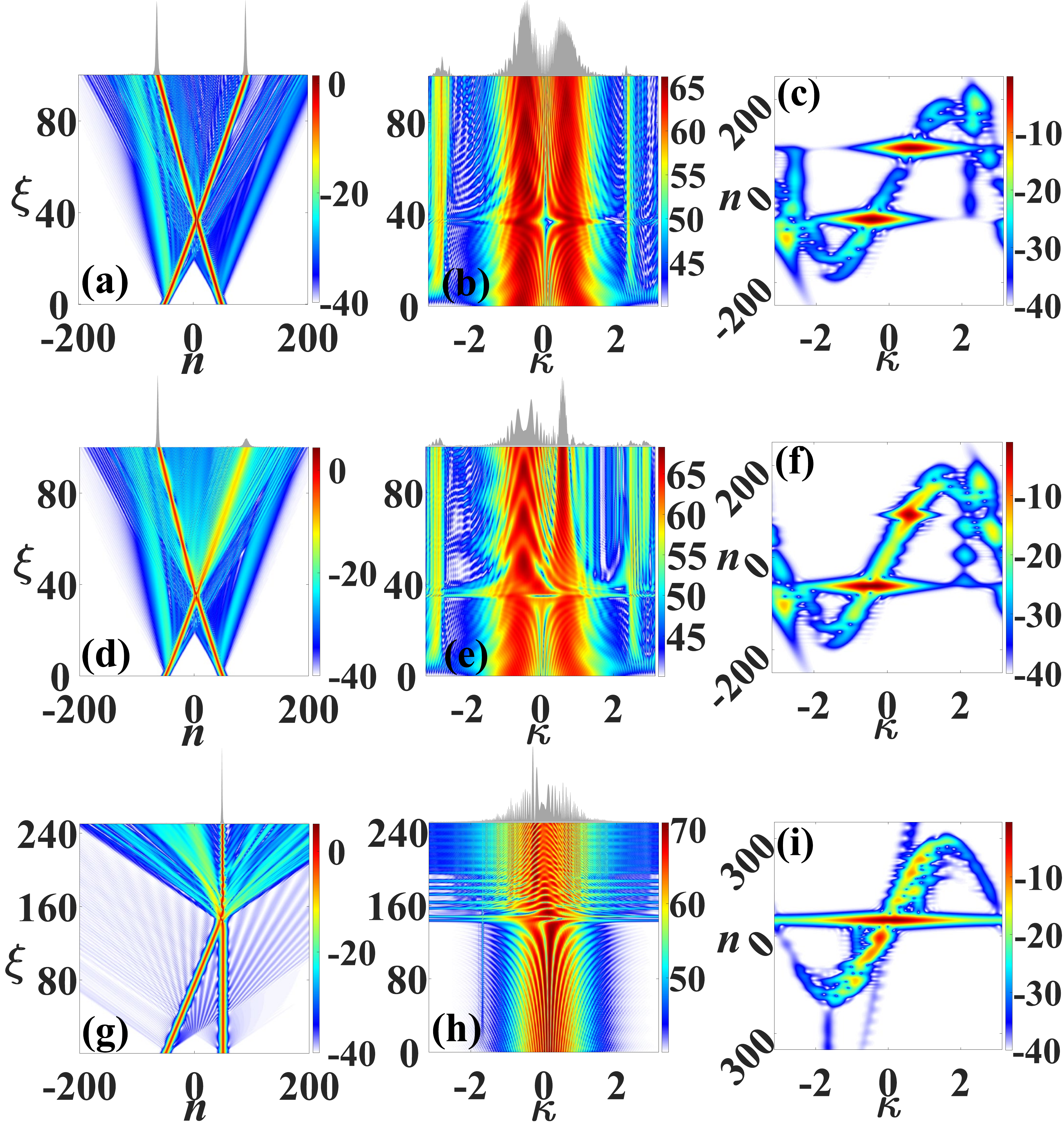}
		\end{center}
		\caption{Collision of two DSs with non-identical wavenumbers ($\abs{k_1} \neq \abs{k_2}$) and equal amplitudes ($A_1=A_2=0.7$). Collision dynamics of DS pair in (a) $n$-space (b) $\kappa$-space and (c) corresponding output spectrogram for the set of parameter $k_1 = 0.9$ and  $k_2 = -0.7$. Collision dynamics of DS pair in (d) $n$-space (e)$\kappa$-space and (f) corresponding output spectrogram for the set of parameter $k_1 = 0.9$ and  $k_2 = -0.6$. Formation of the \textit{fused state} in (g) $n$- space (h) $\kappa$- space and (i) corresponding output spectrogram for the set of parameter $k_1 = 0.3$ and  $k_2 =0$. In all the ceases $n_0=50$ and in the heat maps,  $|\psi_n|^2$ is plotted in log scale along $z$-axis.  
		}
		\label{fig:same_amp}
	\end{figure}
	We observe three distinct collision dynamics when DS pair with different $k$ values (so as widths) are allowed to interact. Depending on the relative value of initial wavenumber $k$, either we have a cross-over state resulting two almost identical DS  or two distinctly different DSs at the output. With suitable $k$ values it is even possible to excite a fused state showing a single soliton at output.  
	To excite crossover states, we consider the wavenumber of one DS to be $k_1 = 0.9$, while taking $k_2 = -0.6$ and $-0.7$, and allow these DS to collide. The entire collision dynamics is illustrated in  Fig.\ref{fig:same_amp} (a)-(f). 
	For the set ($k_1=0.9$ \& $k_2=-0.7$), we observe a negligible  exchange of energy between the DS pair.
	Nothing of interest occurs in $\kappa$-space except the generation of DifRR exhibiting two strong side-bands  (see Fig. \ref{fig:same_amp} (b) and (c)).
	The spectrogram in Fig. \ref{fig:same_amp} (c) indicates the preservation of DS pair and corresponding DifRR recoiled by the Brillouin boundary.
	However strikingly different dynamics is observed for the other set of parameters i.e ($k_1=0.9$ \& $k_2 = -0.6$), where a significant amount of energy is taken away by one DS and the other emerges with a lower amplitude (see Fig. \ref{fig:same_amp} (d)-(f)). We follow this up with a detailed numerical analysis considering a finer variation in the wavenumbers and observe that, the amount of energy transfer between the solitons is highly sensitive to the relative values of $k_1$ and $k_2$ does not follow a pattern in the terms of $k_1$ and $k_2$. Furthermore, the behavior of energy transfer is also dependent upon the launching location $n_0$ of the soliton. Non-identical soliton pair also forms a fused-state exhibiting a single soliton at output. In Fig.\ref{fig:same_amp}(g)-(i) we illustrate the formation of the fused-state while taking $k_1=0.3$ and $k_2=0$.
	The spectrogram in Fig.\ref{fig:same_amp}(i) exhibits the formation of a single fused state with side-lobes.
	Next we consider the most general case where the DSs  have unequal amplitudes ($A_1 \neq A_2$) and initial wavenumbers ($|k_1| \neq |k_2|$).
	We  introduce a scaling parameter $H$ to define one of the DS amplitudes ($A_2 = HA_1$) as function of the other ($A_1$).
	With this substitution we rewrite Eq.\ref{eq:two_sol} as, 
	\begin{align}
		\begin{split}
			\psi(n,0) = A_1 \sech\left[\frac{A_1 (n- n_{1})}{\sqrt{2c \cos(k_1)}}\right]\exp\left[ik_1(n-n_1)\right] \\ 
			+ HA_1 \sech\left[\frac{HA_1(n- n_{2})}{\sqrt{2c \cos(k_2)}}\right]\exp\left[ik_2(n-n_2)\right],
		\end{split}
		\label{eq:neq_sol}
	\end{align}
	while maintaining the combination of $n_{j=1,2}$ and $k_{j=1,2}$ required for the DSs to collide.
	As an initial study, we analytically determine the total energy flowing through WA with a continuous assumption as $E = \int_{-\infty}^{\infty} |\psi(n)|^2 dn$ and obtain  $E=2A_1\sqrt{2c}\left[\cos^{1/2}(k_1)+H\cos^{1/2}(k_2)\right]$.
	We  numerically determine the output energy by exploiting the expression $E = \sum_n |\psi_n|^2$.
	For the set of parameters $c = 1.2$, $A_1 = 0.8$, $H = 0.3$, $k_1 = 0.7$, and $k_2 = -0.8$, we numerically calculate the total energy at output as  $E=2.7885$ which is consistent with    
	 the analytical expression ensuring the conservation of energy.
	\begin{figure}[h]
		\begin{center}
			\includegraphics[width=1\linewidth]{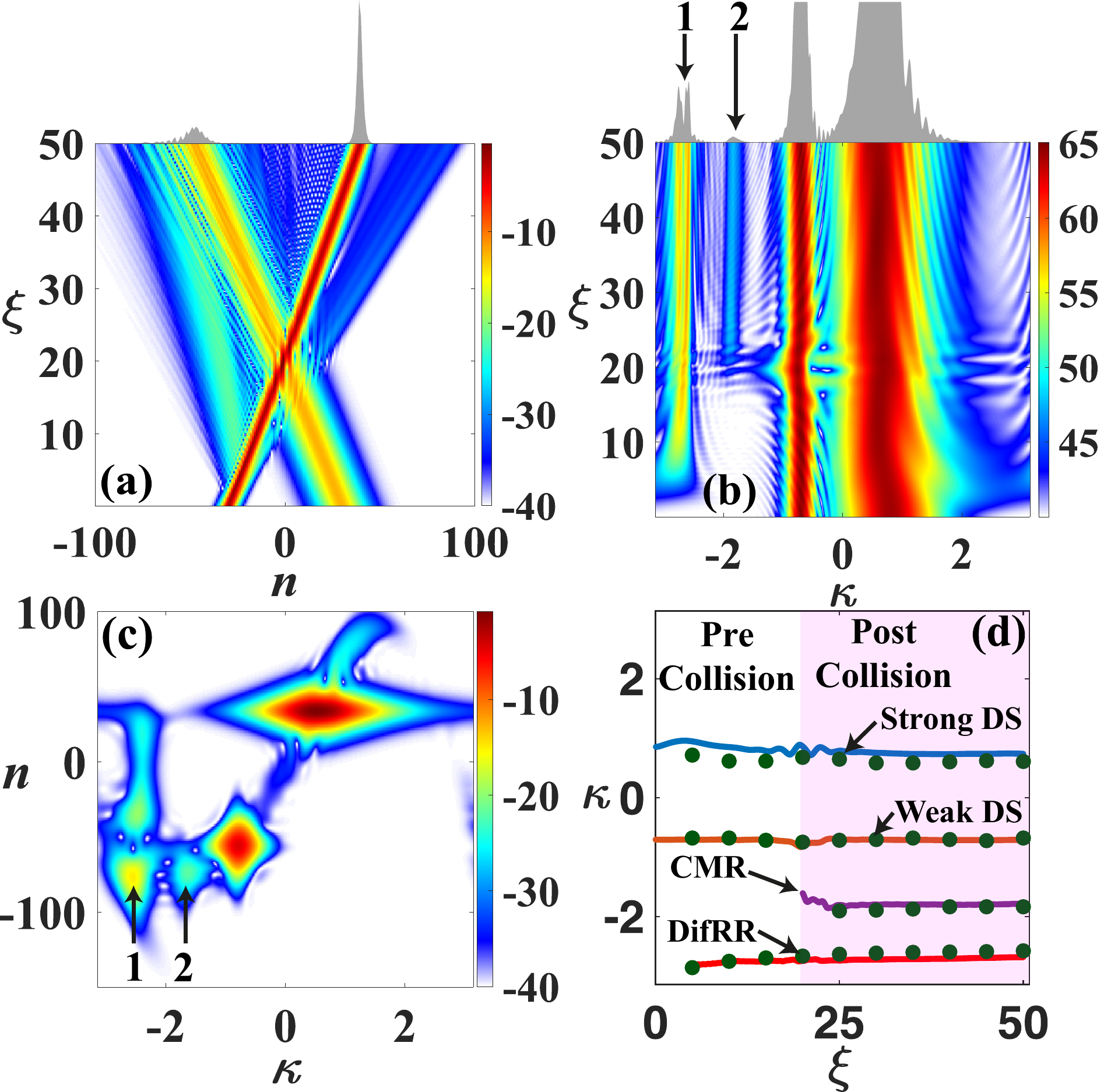}
		\end{center}
		\caption{Collision dynamics of two non-identical DSs illustrated in (a) $n$-space and (b) $\kappa$-space. (c) Spectrogram at output. The the locations of DifRR (1) and secondary radiation (2) are indicated by arrows.  The wave vectors of the two DSs are $k_1 = 0.8$ and $k_2 = -0.7$ respectively, with $A_1=0.8$ and $H=0.3$. 
		Plot (d) describes the evolution of wavenaumbers for four fields, DS pair, DifRR and collision mediated radiation (CMR). The solid lines describe the wavenumbers ($\bar{\kappa}_j$) derived by the \textit{center of mass} technique.  The corresponding numerical results are depicted by solid dots. In all the heat maps  $|\psi_n|^2$ is plotted in log scale along $z$-axis.} 
		\label{fig:NI-WA}
	\end{figure}
	Proceeding ahead, we launch a pair of non-identical DS defined by Eq.\ref{eq:neq_sol} in the proposed WA structure (as shown in Fig.\ref{fig:wa_design}(a)) and depict the collision dynamics in Fig.\ref{fig:NI-WA}. It is evident that the strong DS ($A_1$) almost immediately  emits  DifRR (labeled as 1) whose location  in the $\kappa$-space can be predicted by Eq.\ref{eq:PM}. 
	A secondary  radiation is also originated as a result of the collision between  DS pair. The XFROG diagram shown in Fig.\ref{fig:NI-WA}(c) captures the dynamics with better resolution where we  discern that the secondary  radiation is sandwiched in between primary radiation and  weak DS. To visualize the complete picture see Supplemental Material \cite{movie3}.  The average wavenumber ($\bar {\kappa}_{j}(\xi)$) of the propagating waves can be determined as, \cite{tranControllingDiscreteSoliton2016}, $\bar {\kappa}_{j}(\xi)=\int_{\kappa_i}^{\kappa_f}\kappa|\tilde{\psi}_j(\kappa,\xi)|^2 d\kappa/\int_{\kappa_i}^{\kappa_f}|\tilde{\psi}_j(\kappa,\xi)|^2 d\kappa$, where $\kappa_i$ and $\kappa_f$ determines the range of the distribution of $\tilde{\psi}_j(\kappa)$ in $\kappa$-space.
	In Fig.\ref{fig:NI-WA}(d) we depict the variation of $\bar {\kappa}_{j}(\xi)$ of four different waves (strong DS, weak DS, DifRR and weak secondary radiation) before and after collision. The solid dots represent the corresponding values obtained numerically.  
	The origin of weak secondary radiation can be understood by applying the concept of \textit{blocker soliton} \textcolor{red} {\cite{Meier:05}}. The  \textit{blocker soliton} is strongly localized wave that behave as a reflector to a weak signal beam. In Fig. \ref{fig13} we  demonstrate the interaction between  a  \textit{blocker soliton} and  weak signal beam. The \textit{blocker soliton} doesn't radiation any DifRR as the transverse wavenumber is zero. However, a part of the  weak DS (signal beam) is deflected by   \textit{blocker soliton} and results in a radiation at $\kappa$-space. In Fig \ref{fig13} (a),(b) we demonstrate the interaction dynamics of solitons in $n$ and $\kappa$-space where the radiation due to the deflection is evident. The corresponding spectrograme is shown in Fig \ref{fig13} (c). The exact phase matching condition for such radiation is found to be complicated and  may require a more detailed analysis.

	\begin{figure}[h]
		\begin{center}
			\includegraphics[width=1.0\linewidth]{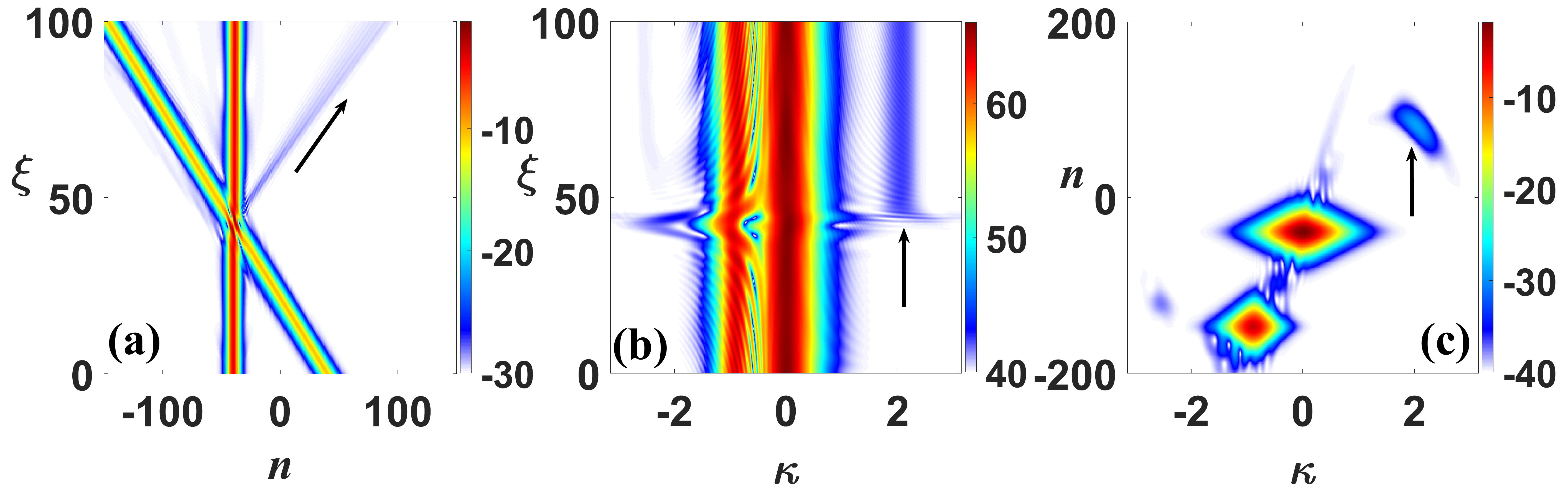}
		\end{center}
		\caption{(a) Collision dynamics between a strong (blocker) and weak DS where the weak soliton is deflected (shown by arrow). (b) Dynamics shown in $\kappa$-space where the collision mediated radiation is indicated by the arrow. (c) Corresponding spectrograme also shoing the radiation patch (shown by arrow). The parameters taken for the simulation are, $A_1=0.6$, $H=0.5$, $k_1=0$, $k_2=-0.9$ and $n_1=n_2=40$ .}
		\label{fig13}
	\end{figure}

	\section{Conclusion}
	A spatial discrete soliton emits diffractive resonance radiation when launched with an initial wavenumber in a semi-infinite uniform waveguide array. 
	The situation becomes even interesting when two such solitons are allowed to collide with each other by inserting a \textit{push} through initial wavenumber resulting an oblique incidence. 
	We  numerically  investigate the complex collision dynamics of varied types of spatial DS pairs, staring from the simplest case of identical DS pair, to a more generalized combination of soliton pairs.
	The collision of the identical DS pair,  either forms $(i)$ a fused \textit{breather} state or $(ii)$ a cross-over state (followed by an elastic like collision) depending on the relative values of initial amplitudes ($A_0$) and wave number ($k_0$) of the solitons. For relatively high $A_0$ and low $k_0$ results in a \textit{stop band} where the two solitons propagate without any interaction by forming a bound-state. We map this entire process by depicting a phase plot in  ($k_0$-$A_0$) parameter space where different regions are indicated by color bands. 
    We  establish a semi-analytical approach using variational method to understand the collision mechanism between DS. DSs that extend over few waveguide channels allow us to approximate the discrete transverse variable with a continuous one and we rewrite the DNLSE to a perturbed coupled NLSE. By adopting a proper Lagrangian density and suitable  \textit{ans\"{a}tz} we derive two sets of four coupled ordinary differential equations that describe the evolution of soliton parameters (amplitude, position, wavenumber and phase) under collision. The variational results reveal interesting facts on collision dynamics and corroborate well with full numerical simulation. 
	The relative phase difference measurement also results in very different observations for \textit{breather} formation and cross-over state.
	When for a crossover state, the relative phase changes  monotonically with negative value, the \textit{breather} formation attributes a positive increment in phase. 
	It is noticed that the phase changes abruptly at collision points.
	We investigate the role of initial phase detuning between the colliding DS. It is observed that the \textit{breather} formation is ceased for two out of phase ($\Delta \Phi=\pi$) DSs. We notice that, periodic energy exchange takes place between the DSs when initial phase is detuned in the range $0<\Delta \Phi<\pi$.
	 We finally analyze the interaction for two non-identical soliton and find that along with  the usual \textit{breather} state, two different cross-over states are formed where degree of energy exchange between solitons differs significantly.   
	By extending this study to a more generalized condition of both DSs having different amplitudes and wavenumbers, we observe the formation of a collision mediated secondary radiation which is originated due to  reflection of the weaker soliton from the edge of the strong \textit {blocker soliton}.  
	In conclusion we can say, collision dynamics of spatial DSs in discrete media yields a very diverse range of interesting phenomena and in our study we try to shed light on few of them. Our results should be useful to the research community doing experiments on DSs in WA.
	
	\section*{ACKNOWLEDGMENT}
	A.P.L. acknowledges University Grants Commission, India for support through Junior Research Fellowship in Sciences,	Humanities and Social Sciences (ID 515364)
	
		
	\bibliography{reference.bib}

\begin{thebibliography}{69}
\expandafter\ifx\csname natexlab\endcsname\relax\def\natexlab#1{#1}\fi
\expandafter\ifx\csname bibnamefont\endcsname\relax
  \def\bibnamefont#1{#1}\fi
\expandafter\ifx\csname bibfnamefont\endcsname\relax
  \def\bibfnamefont#1{#1}\fi
\expandafter\ifx\csname citenamefont\endcsname\relax
  \def\citenamefont#1{#1}\fi
\expandafter\ifx\csname url\endcsname\relax
  \def\url#1{\texttt{#1}}\fi
\expandafter\ifx\csname urlprefix\endcsname\relax\def\urlprefix{URL }\fi
\providecommand{\bibinfo}[2]{#2}
\providecommand{\eprint}[2][]{\url{#2}}

\bibitem[{\citenamefont{Miller}(1954)}]{millerCoupledWaveTheory1954}
\bibinfo{author}{\bibfnamefont{S.~E.} \bibnamefont{Miller}},
  \bibinfo{journal}{Bell System Technical Journal}
  \textbf{\bibinfo{volume}{33}}, \bibinfo{pages}{661} (\bibinfo{year}{1954}).

\bibitem[{\citenamefont{Yariv}(1973)}]{yarivCoupledmodeTheoryGuidedwave1973}
\bibinfo{author}{\bibfnamefont{A.}~\bibnamefont{Yariv}}, \bibinfo{journal}{IEEE
  Journal of Quantum Electronics} \textbf{\bibinfo{volume}{9}},
  \bibinfo{pages}{919} (\bibinfo{year}{1973}).

\bibitem[{\citenamefont{Das et~al.}(1987)\citenamefont{Das, Chen, and
  Bhattacharya}}]{dasNonlinearEffectsCoplanar1987}
\bibinfo{author}{\bibfnamefont{U.}~\bibnamefont{Das}},
  \bibinfo{author}{\bibfnamefont{Y.}~\bibnamefont{Chen}}, \bibnamefont{and}
  \bibinfo{author}{\bibfnamefont{P.}~\bibnamefont{Bhattacharya}},
  \bibinfo{journal}{Applied Physics Letters} \textbf{\bibinfo{volume}{51}},
  \bibinfo{pages}{1679} (\bibinfo{year}{1987}).

\bibitem[{\citenamefont{Haus and
  {Molter-Orr}}(1983)}]{hausCoupledMultipleWaveguide1983}
\bibinfo{author}{\bibfnamefont{H.}~\bibnamefont{Haus}} \bibnamefont{and}
  \bibinfo{author}{\bibfnamefont{L.}~\bibnamefont{{Molter-Orr}}},
  \bibinfo{journal}{IEEE Journal of Quantum Electronics}
  \textbf{\bibinfo{volume}{19}}, \bibinfo{pages}{840} (\bibinfo{year}{1983}).

\bibitem[{\citenamefont{Finlayson and
  Stegeman}(1990)}]{finlaysonSpatialSwitchingInstabilities1990}
\bibinfo{author}{\bibfnamefont{N.}~\bibnamefont{Finlayson}} \bibnamefont{and}
  \bibinfo{author}{\bibfnamefont{G.~I.} \bibnamefont{Stegeman}},
  \bibinfo{journal}{Applied Physics Letters} \textbf{\bibinfo{volume}{56}},
  \bibinfo{pages}{2276} (\bibinfo{year}{1990}).

\bibitem[{\citenamefont{Christodoulides and
  Joseph}(1988)}]{christodoulidesDiscreteSelffocusingNonlinear1988a}
\bibinfo{author}{\bibfnamefont{D.~N.} \bibnamefont{Christodoulides}}
  \bibnamefont{and} \bibinfo{author}{\bibfnamefont{R.~I.}
  \bibnamefont{Joseph}}, \bibinfo{journal}{Optics Letters}
  \textbf{\bibinfo{volume}{13}}, \bibinfo{pages}{794} (\bibinfo{year}{1988}).

\bibitem[{\citenamefont{Sipe and
  Winful}(1988)}]{sipeNonlinearSchrodingerSolitons1988a}
\bibinfo{author}{\bibfnamefont{J.~E.} \bibnamefont{Sipe}} \bibnamefont{and}
  \bibinfo{author}{\bibfnamefont{H.~G.} \bibnamefont{Winful}},
  \bibinfo{journal}{Optics Letters} \textbf{\bibinfo{volume}{13}},
  \bibinfo{pages}{132} (\bibinfo{year}{1988}).

\bibitem[{\citenamefont{Kivshar and
  Campbell}(1993)}]{kivsharPeierlsNabarroPotentialBarrier1993}
\bibinfo{author}{\bibfnamefont{Y.~S.} \bibnamefont{Kivshar}} \bibnamefont{and}
  \bibinfo{author}{\bibfnamefont{D.~K.} \bibnamefont{Campbell}},
  \bibinfo{journal}{Physical Review E} \textbf{\bibinfo{volume}{48}},
  \bibinfo{pages}{3077} (\bibinfo{year}{1993}).

\bibitem[{\citenamefont{Kivshar}(1993)}]{kivsharSelflocalizationArraysDefocusing1993}
\bibinfo{author}{\bibfnamefont{Y.~S.} \bibnamefont{Kivshar}},
  \bibinfo{journal}{Optics Letters} \textbf{\bibinfo{volume}{18}},
  \bibinfo{pages}{1147} (\bibinfo{year}{1993}).

\bibitem[{\citenamefont{Pertsch et~al.}(2002)\citenamefont{Pertsch, Zentgraf,
  Peschel, Br{\"a}uer, and
  Lederer}}]{pertschAnomalousRefractionDiffraction2002a}
\bibinfo{author}{\bibfnamefont{T.}~\bibnamefont{Pertsch}},
  \bibinfo{author}{\bibfnamefont{T.}~\bibnamefont{Zentgraf}},
  \bibinfo{author}{\bibfnamefont{U.}~\bibnamefont{Peschel}},
  \bibinfo{author}{\bibfnamefont{A.}~\bibnamefont{Br{\"a}uer}},
  \bibnamefont{and} \bibinfo{author}{\bibfnamefont{F.}~\bibnamefont{Lederer}},
  \bibinfo{journal}{Physical Review Letters} \textbf{\bibinfo{volume}{88}},
  \bibinfo{pages}{093901} (\bibinfo{year}{2002}).

\bibitem[{\citenamefont{Eisenberg et~al.}(1998)\citenamefont{Eisenberg,
  Silberberg, Morandotti, Boyd, and
  Aitchison}}]{eisenbergDiscreteSpatialOptical1998}
\bibinfo{author}{\bibfnamefont{H.~S.} \bibnamefont{Eisenberg}},
  \bibinfo{author}{\bibfnamefont{Y.}~\bibnamefont{Silberberg}},
  \bibinfo{author}{\bibfnamefont{R.}~\bibnamefont{Morandotti}},
  \bibinfo{author}{\bibfnamefont{A.~R.} \bibnamefont{Boyd}}, \bibnamefont{and}
  \bibinfo{author}{\bibfnamefont{J.~S.} \bibnamefont{Aitchison}},
  \bibinfo{journal}{Physical Review Letters} \textbf{\bibinfo{volume}{81}},
  \bibinfo{pages}{3383} (\bibinfo{year}{1998}).

\bibitem[{\citenamefont{Eisenberg et~al.}(2002)\citenamefont{Eisenberg,
  Morandotti, Silberberg, Arnold, Pennelli, and
  Aitchison}}]{eisenbergOpticalDiscreteSolitons2002a}
\bibinfo{author}{\bibfnamefont{H.~S.} \bibnamefont{Eisenberg}},
  \bibinfo{author}{\bibfnamefont{R.}~\bibnamefont{Morandotti}},
  \bibinfo{author}{\bibfnamefont{Y.}~\bibnamefont{Silberberg}},
  \bibinfo{author}{\bibfnamefont{J.~M.} \bibnamefont{Arnold}},
  \bibinfo{author}{\bibfnamefont{G.}~\bibnamefont{Pennelli}}, \bibnamefont{and}
  \bibinfo{author}{\bibfnamefont{J.~S.} \bibnamefont{Aitchison}},
  \bibinfo{journal}{Journal of the Optical Society of America B}
  \textbf{\bibinfo{volume}{19}}, \bibinfo{pages}{2938} (\bibinfo{year}{2002}).

\bibitem[{\citenamefont{Peschel et~al.}(2002)\citenamefont{Peschel, Morandotti,
  Arnold, Aitchison, Eisenberg, Silberberg, Pertsch, and
  Lederer}}]{OpticalDiscreteSolitons2002}
\bibinfo{author}{\bibfnamefont{U.}~\bibnamefont{Peschel}},
  \bibinfo{author}{\bibfnamefont{R.}~\bibnamefont{Morandotti}},
  \bibinfo{author}{\bibfnamefont{J.~M.} \bibnamefont{Arnold}},
  \bibinfo{author}{\bibfnamefont{J.~S.} \bibnamefont{Aitchison}},
  \bibinfo{author}{\bibfnamefont{H.~S.} \bibnamefont{Eisenberg}},
  \bibinfo{author}{\bibfnamefont{Y.}~\bibnamefont{Silberberg}},
  \bibinfo{author}{\bibfnamefont{T.}~\bibnamefont{Pertsch}}, \bibnamefont{and}
  \bibinfo{author}{\bibfnamefont{F.}~\bibnamefont{Lederer}},
  \bibinfo{journal}{Journal of the Optical Society of America B}
  \textbf{\bibinfo{volume}{19}}, \bibinfo{pages}{2637} (\bibinfo{year}{2002}).

\bibitem[{\citenamefont{Aceves et~al.}(1996)\citenamefont{Aceves, De~Angelis,
  Peschel, Muschall, Lederer, Trillo, and
  Wabnitz}}]{acevesDiscreteSelftrappingSoliton1996}
\bibinfo{author}{\bibfnamefont{A.~B.} \bibnamefont{Aceves}},
  \bibinfo{author}{\bibfnamefont{C.}~\bibnamefont{De~Angelis}},
  \bibinfo{author}{\bibfnamefont{T.}~\bibnamefont{Peschel}},
  \bibinfo{author}{\bibfnamefont{R.}~\bibnamefont{Muschall}},
  \bibinfo{author}{\bibfnamefont{F.}~\bibnamefont{Lederer}},
  \bibinfo{author}{\bibfnamefont{S.}~\bibnamefont{Trillo}}, \bibnamefont{and}
  \bibinfo{author}{\bibfnamefont{S.}~\bibnamefont{Wabnitz}},
  \bibinfo{journal}{Physical Review E} \textbf{\bibinfo{volume}{53}},
  \bibinfo{pages}{1172} (\bibinfo{year}{1996}).

\bibitem[{\citenamefont{Cai et~al.}(1994)\citenamefont{Cai, Bishop, and
  {Gr{\o}nbech-Jensen}}}]{caiLocalizedStatesDiscrete1994}
\bibinfo{author}{\bibfnamefont{D.}~\bibnamefont{Cai}},
  \bibinfo{author}{\bibfnamefont{A.~R.} \bibnamefont{Bishop}},
  \bibnamefont{and}
  \bibinfo{author}{\bibfnamefont{N.}~\bibnamefont{{Gr{\o}nbech-Jensen}}},
  \bibinfo{journal}{Physical Review Letters} \textbf{\bibinfo{volume}{72}},
  \bibinfo{pages}{591} (\bibinfo{year}{1994}).

\bibitem[{\citenamefont{Morandotti
  et~al.}(1999{\natexlab{a}})\citenamefont{Morandotti, Peschel, Aitchison,
  Eisenberg, and Silberberg}}]{morandottiDynamicsDiscreteSolitons1999}
\bibinfo{author}{\bibfnamefont{R.}~\bibnamefont{Morandotti}},
  \bibinfo{author}{\bibfnamefont{U.}~\bibnamefont{Peschel}},
  \bibinfo{author}{\bibfnamefont{J.~S.} \bibnamefont{Aitchison}},
  \bibinfo{author}{\bibfnamefont{H.~S.} \bibnamefont{Eisenberg}},
  \bibnamefont{and}
  \bibinfo{author}{\bibfnamefont{Y.}~\bibnamefont{Silberberg}},
  \bibinfo{journal}{Phys. Rev. Lett.} \textbf{\bibinfo{volume}{83}},
  \bibinfo{pages}{2726} (\bibinfo{year}{1999}{\natexlab{a}}),
  \urlprefix\url{https://link.aps.org/doi/10.1103/PhysRevLett.83.2726}.

\bibitem[{\citenamefont{Lahini et~al.}(2008)\citenamefont{Lahini, Avidan,
  Pozzi, Sorel, Morandotti, Christodoulides, and
  Silberberg}}]{lahiniAndersonLocalizationNonlinearity2008}
\bibinfo{author}{\bibfnamefont{Y.}~\bibnamefont{Lahini}},
  \bibinfo{author}{\bibfnamefont{A.}~\bibnamefont{Avidan}},
  \bibinfo{author}{\bibfnamefont{F.}~\bibnamefont{Pozzi}},
  \bibinfo{author}{\bibfnamefont{M.}~\bibnamefont{Sorel}},
  \bibinfo{author}{\bibfnamefont{R.}~\bibnamefont{Morandotti}},
  \bibinfo{author}{\bibfnamefont{D.~N.} \bibnamefont{Christodoulides}},
  \bibnamefont{and}
  \bibinfo{author}{\bibfnamefont{Y.}~\bibnamefont{Silberberg}},
  \bibinfo{journal}{Physical Review Letters} \textbf{\bibinfo{volume}{100}},
  \bibinfo{pages}{013906} (\bibinfo{year}{2008}).

\bibitem[{\citenamefont{Martin et~al.}(2011)\citenamefont{Martin, Giuseppe,
  {Perez-Leija}, Keil, Dreisow, Heinrich, Nolte, Szameit, Abouraddy,
  Christodoulides et~al.}}]{martinAndersonLocalizationOptical2011}
\bibinfo{author}{\bibfnamefont{L.}~\bibnamefont{Martin}},
  \bibinfo{author}{\bibfnamefont{G.~D.} \bibnamefont{Giuseppe}},
  \bibinfo{author}{\bibfnamefont{A.}~\bibnamefont{{Perez-Leija}}},
  \bibinfo{author}{\bibfnamefont{R.}~\bibnamefont{Keil}},
  \bibinfo{author}{\bibfnamefont{F.}~\bibnamefont{Dreisow}},
  \bibinfo{author}{\bibfnamefont{M.}~\bibnamefont{Heinrich}},
  \bibinfo{author}{\bibfnamefont{S.}~\bibnamefont{Nolte}},
  \bibinfo{author}{\bibfnamefont{A.}~\bibnamefont{Szameit}},
  \bibinfo{author}{\bibfnamefont{A.~F.} \bibnamefont{Abouraddy}},
  \bibinfo{author}{\bibfnamefont{D.~N.} \bibnamefont{Christodoulides}},
  \bibnamefont{et~al.}, p.~\bibinfo{pages}{11} (\bibinfo{year}{2011}).

\bibitem[{\citenamefont{Morandotti
  et~al.}(1999{\natexlab{b}})\citenamefont{Morandotti, Peschel, Aitchison,
  Eisenberg, and Silberberg}}]{morandottiExperimentalObservationLinear1999}
\bibinfo{author}{\bibfnamefont{R.}~\bibnamefont{Morandotti}},
  \bibinfo{author}{\bibfnamefont{U.}~\bibnamefont{Peschel}},
  \bibinfo{author}{\bibfnamefont{J.~S.} \bibnamefont{Aitchison}},
  \bibinfo{author}{\bibfnamefont{H.~S.} \bibnamefont{Eisenberg}},
  \bibnamefont{and}
  \bibinfo{author}{\bibfnamefont{Y.}~\bibnamefont{Silberberg}},
  \bibinfo{journal}{Physical Review Letters} \textbf{\bibinfo{volume}{83}},
  \bibinfo{pages}{4756} (\bibinfo{year}{1999}{\natexlab{b}}).

\bibitem[{\citenamefont{Pertsch et~al.}(1999)\citenamefont{Pertsch, Dannberg,
  Elflein, Br{\"a}uer, and Lederer}}]{pertschOpticalBlochOscillations1999}
\bibinfo{author}{\bibfnamefont{T.}~\bibnamefont{Pertsch}},
  \bibinfo{author}{\bibfnamefont{P.}~\bibnamefont{Dannberg}},
  \bibinfo{author}{\bibfnamefont{W.}~\bibnamefont{Elflein}},
  \bibinfo{author}{\bibfnamefont{A.}~\bibnamefont{Br{\"a}uer}},
  \bibnamefont{and} \bibinfo{author}{\bibfnamefont{F.}~\bibnamefont{Lederer}},
  \bibinfo{journal}{Physical Review Letters} \textbf{\bibinfo{volume}{83}},
  \bibinfo{pages}{4752} (\bibinfo{year}{1999}).

\bibitem[{\citenamefont{Breid et~al.}(2006)\citenamefont{Breid, Witthaut, and
  Korsch}}]{breidBlochZenerOscillations2006}
\bibinfo{author}{\bibfnamefont{B.~M.} \bibnamefont{Breid}},
  \bibinfo{author}{\bibfnamefont{D.}~\bibnamefont{Witthaut}}, \bibnamefont{and}
  \bibinfo{author}{\bibfnamefont{H.~J.} \bibnamefont{Korsch}},
  \bibinfo{journal}{New Journal of Physics} \textbf{\bibinfo{volume}{8}},
  \bibinfo{pages}{110} (\bibinfo{year}{2006}).

\bibitem[{\citenamefont{Dreisow et~al.}(2009)\citenamefont{Dreisow, Szameit,
  Heinrich, Pertsch, Nolte, T\"unnermann, and
  Longhi}}]{dreisowBlochZenerOscillationsBinary2009}
\bibinfo{author}{\bibfnamefont{F.}~\bibnamefont{Dreisow}},
  \bibinfo{author}{\bibfnamefont{A.}~\bibnamefont{Szameit}},
  \bibinfo{author}{\bibfnamefont{M.}~\bibnamefont{Heinrich}},
  \bibinfo{author}{\bibfnamefont{T.}~\bibnamefont{Pertsch}},
  \bibinfo{author}{\bibfnamefont{S.}~\bibnamefont{Nolte}},
  \bibinfo{author}{\bibfnamefont{A.}~\bibnamefont{T\"unnermann}},
  \bibnamefont{and} \bibinfo{author}{\bibfnamefont{S.}~\bibnamefont{Longhi}},
  \bibinfo{journal}{Phys. Rev. Lett.} \textbf{\bibinfo{volume}{102}},
  \bibinfo{pages}{076802} (\bibinfo{year}{2009}),
  \urlprefix\url{https://link.aps.org/doi/10.1103/PhysRevLett.102.076802}.

\bibitem[{\citenamefont{Longhi et~al.}(2006)\citenamefont{Longhi, Marangoni,
  Lobino, Ramponi, Laporta, Cianci, and
  Foglietti}}]{longhiObservationDynamicLocalization2006}
\bibinfo{author}{\bibfnamefont{S.}~\bibnamefont{Longhi}},
  \bibinfo{author}{\bibfnamefont{M.}~\bibnamefont{Marangoni}},
  \bibinfo{author}{\bibfnamefont{M.}~\bibnamefont{Lobino}},
  \bibinfo{author}{\bibfnamefont{R.}~\bibnamefont{Ramponi}},
  \bibinfo{author}{\bibfnamefont{P.}~\bibnamefont{Laporta}},
  \bibinfo{author}{\bibfnamefont{E.}~\bibnamefont{Cianci}}, \bibnamefont{and}
  \bibinfo{author}{\bibfnamefont{V.}~\bibnamefont{Foglietti}},
  \bibinfo{journal}{Physical Review Letters} \textbf{\bibinfo{volume}{96}},
  \bibinfo{pages}{243901} (\bibinfo{year}{2006}).

\bibitem[{\citenamefont{Hizanidis et~al.}(2008)\citenamefont{Hizanidis,
  Kominis, and Efremidis}}]{hizanidisInterlacedLinearnonlinearOptical2008}
\bibinfo{author}{\bibfnamefont{K.}~\bibnamefont{Hizanidis}},
  \bibinfo{author}{\bibfnamefont{Y.}~\bibnamefont{Kominis}}, \bibnamefont{and}
  \bibinfo{author}{\bibfnamefont{N.~K.} \bibnamefont{Efremidis}},
  \bibinfo{journal}{Optics Express} \textbf{\bibinfo{volume}{16}},
  \bibinfo{pages}{18296} (\bibinfo{year}{2008}).

\bibitem[{\citenamefont{Minardi et~al.}(2010)\citenamefont{Minardi,
  Eilenberger, Kartashov, Szameit, R\"opke, Kobelke, Schuster, Bartelt, Nolte,
  Torner et~al.}}]{minardiThreeDimensionalLightBullets2010}
\bibinfo{author}{\bibfnamefont{S.}~\bibnamefont{Minardi}},
  \bibinfo{author}{\bibfnamefont{F.}~\bibnamefont{Eilenberger}},
  \bibinfo{author}{\bibfnamefont{Y.~V.} \bibnamefont{Kartashov}},
  \bibinfo{author}{\bibfnamefont{A.}~\bibnamefont{Szameit}},
  \bibinfo{author}{\bibfnamefont{U.}~\bibnamefont{R\"opke}},
  \bibinfo{author}{\bibfnamefont{J.}~\bibnamefont{Kobelke}},
  \bibinfo{author}{\bibfnamefont{K.}~\bibnamefont{Schuster}},
  \bibinfo{author}{\bibfnamefont{H.}~\bibnamefont{Bartelt}},
  \bibinfo{author}{\bibfnamefont{S.}~\bibnamefont{Nolte}},
  \bibinfo{author}{\bibfnamefont{L.}~\bibnamefont{Torner}},
  \bibnamefont{et~al.}, \bibinfo{journal}{Phys. Rev. Lett.}
  \textbf{\bibinfo{volume}{105}}, \bibinfo{pages}{263901}
  (\bibinfo{year}{2010}),
  \urlprefix\url{https://link.aps.org/doi/10.1103/PhysRevLett.105.263901}.

\bibitem[{\citenamefont{Mihalache
  et~al.}(2008{\natexlab{a}})\citenamefont{Mihalache, Mazilu, Lederer, and
  Kivshar}}]{mihalacheSpatiotemporalDissipativeSolitons2008a}
\bibinfo{author}{\bibfnamefont{D.}~\bibnamefont{Mihalache}},
  \bibinfo{author}{\bibfnamefont{D.}~\bibnamefont{Mazilu}},
  \bibinfo{author}{\bibfnamefont{F.}~\bibnamefont{Lederer}}, \bibnamefont{and}
  \bibinfo{author}{\bibfnamefont{Y.~S.} \bibnamefont{Kivshar}},
  \bibinfo{journal}{Phys. Rev. E} \textbf{\bibinfo{volume}{78}},
  \bibinfo{pages}{056602} (\bibinfo{year}{2008}{\natexlab{a}}),
  \urlprefix\url{https://link.aps.org/doi/10.1103/PhysRevE.78.056602}.

\bibitem[{\citenamefont{Mihalache
  et~al.}(2008{\natexlab{b}})\citenamefont{Mihalache, Mazilu, Lederer, and
  Kivshar}}]{mihalacheSpatiotemporalSurfaceGinzburgLandau2008}
\bibinfo{author}{\bibfnamefont{D.}~\bibnamefont{Mihalache}},
  \bibinfo{author}{\bibfnamefont{D.}~\bibnamefont{Mazilu}},
  \bibinfo{author}{\bibfnamefont{F.}~\bibnamefont{Lederer}}, \bibnamefont{and}
  \bibinfo{author}{\bibfnamefont{Y.~S.} \bibnamefont{Kivshar}},
  \bibinfo{journal}{Phys. Rev. A} \textbf{\bibinfo{volume}{77}},
  \bibinfo{pages}{043828} (\bibinfo{year}{2008}{\natexlab{b}}),
  \urlprefix\url{https://link.aps.org/doi/10.1103/PhysRevA.77.043828}.

\bibitem[{\citenamefont{Longhi}(2010{\natexlab{a}})}]{longhiPhotonicAnalogZitterbewegung2010}
\bibinfo{author}{\bibfnamefont{S.}~\bibnamefont{Longhi}},
  \bibinfo{journal}{Optics Letters} \textbf{\bibinfo{volume}{35}},
  \bibinfo{pages}{235} (\bibinfo{year}{2010}{\natexlab{a}}).

\bibitem[{\citenamefont{Longhi}(2010{\natexlab{b}})}]{longhiKleinTunnelingBinary2010}
\bibinfo{author}{\bibfnamefont{S.}~\bibnamefont{Longhi}},
  \bibinfo{journal}{Phys. Rev. B} \textbf{\bibinfo{volume}{81}},
  \bibinfo{pages}{075102} (\bibinfo{year}{2010}{\natexlab{b}}),
  \urlprefix\url{https://link.aps.org/doi/10.1103/PhysRevB.81.075102}.

\bibitem[{\citenamefont{Keil et~al.}(2011)\citenamefont{Keil, {Perez-Leija},
  Dreisow, Heinrich, {Moya-Cessa}, Nolte, Christodoulides, and
  Szameit}}]{keilClassicalAnalogueDisplaced2011}
\bibinfo{author}{\bibfnamefont{R.}~\bibnamefont{Keil}},
  \bibinfo{author}{\bibfnamefont{A.}~\bibnamefont{{Perez-Leija}}},
  \bibinfo{author}{\bibfnamefont{F.}~\bibnamefont{Dreisow}},
  \bibinfo{author}{\bibfnamefont{M.}~\bibnamefont{Heinrich}},
  \bibinfo{author}{\bibfnamefont{H.}~\bibnamefont{{Moya-Cessa}}},
  \bibinfo{author}{\bibfnamefont{S.}~\bibnamefont{Nolte}},
  \bibinfo{author}{\bibfnamefont{D.~N.} \bibnamefont{Christodoulides}},
  \bibnamefont{and} \bibinfo{author}{\bibfnamefont{A.}~\bibnamefont{Szameit}},
  \bibinfo{journal}{Physical Review Letters} \textbf{\bibinfo{volume}{107}},
  \bibinfo{pages}{103601} (\bibinfo{year}{2011}).

\bibitem[{\citenamefont{Marini et~al.}(2014)\citenamefont{Marini, Longhi, and
  Biancalana}}]{mariniOpticalSimulationNeutrino2014}
\bibinfo{author}{\bibfnamefont{A.}~\bibnamefont{Marini}},
  \bibinfo{author}{\bibfnamefont{S.}~\bibnamefont{Longhi}}, \bibnamefont{and}
  \bibinfo{author}{\bibfnamefont{F.}~\bibnamefont{Biancalana}},
  \bibinfo{journal}{Physical Review Letters} \textbf{\bibinfo{volume}{113}},
  \bibinfo{pages}{150401} (\bibinfo{year}{2014}).

\bibitem[{\citenamefont{Tran et~al.}(2014)\citenamefont{Tran, Longhi, and
  Biancalana}}]{tranOpticalAnalogueRelativistic2014}
\bibinfo{author}{\bibfnamefont{T.~X.} \bibnamefont{Tran}},
  \bibinfo{author}{\bibfnamefont{S.}~\bibnamefont{Longhi}}, \bibnamefont{and}
  \bibinfo{author}{\bibfnamefont{F.}~\bibnamefont{Biancalana}},
  \bibinfo{journal}{Annals of Physics} \textbf{\bibinfo{volume}{340}},
  \bibinfo{pages}{179} (\bibinfo{year}{2014}).

\bibitem[{\citenamefont{Williams and
  Kutz}(2012)}]{williamsGeneratingRoutingLightbullets2012a}
\bibinfo{author}{\bibfnamefont{M.~O.} \bibnamefont{Williams}} \bibnamefont{and}
  \bibinfo{author}{\bibfnamefont{J.~N.} \bibnamefont{Kutz}},
  \bibinfo{journal}{Optical and Quantum Electronics}
  \textbf{\bibinfo{volume}{44}}, \bibinfo{pages}{247} (\bibinfo{year}{2012}).

\bibitem[{\citenamefont{Zhang et~al.}(2017)\citenamefont{Zhang, Yuan, Xu, and
  Ye}}]{zhangManipulatingDiscreteSolitons2017}
\bibinfo{author}{\bibfnamefont{X.}~\bibnamefont{Zhang}},
  \bibinfo{author}{\bibfnamefont{X.}~\bibnamefont{Yuan}},
  \bibinfo{author}{\bibfnamefont{W.}~\bibnamefont{Xu}}, \bibnamefont{and}
  \bibinfo{author}{\bibfnamefont{W.}~\bibnamefont{Ye}},
  \bibinfo{journal}{Optics Express} \textbf{\bibinfo{volume}{25}},
  \bibinfo{pages}{31204} (\bibinfo{year}{2017}).

\bibitem[{\citenamefont{Block et~al.}(2014)\citenamefont{Block, Etrich,
  Limboeck, Bleckmann, Soergel, Rockstuhl, and
  Linden}}]{blockBlochOscillationsPlasmonic2014}
\bibinfo{author}{\bibfnamefont{A.}~\bibnamefont{Block}},
  \bibinfo{author}{\bibfnamefont{C.}~\bibnamefont{Etrich}},
  \bibinfo{author}{\bibfnamefont{T.}~\bibnamefont{Limboeck}},
  \bibinfo{author}{\bibfnamefont{F.}~\bibnamefont{Bleckmann}},
  \bibinfo{author}{\bibfnamefont{E.}~\bibnamefont{Soergel}},
  \bibinfo{author}{\bibfnamefont{C.}~\bibnamefont{Rockstuhl}},
  \bibnamefont{and} \bibinfo{author}{\bibfnamefont{S.}~\bibnamefont{Linden}},
  \bibinfo{journal}{Nature Communications} \textbf{\bibinfo{volume}{5}},
  \bibinfo{pages}{3843} (\bibinfo{year}{2014}).

\bibitem[{\citenamefont{Pezzi et~al.}(2019)\citenamefont{Pezzi, De~Sio, Veltri,
  Cunningham, De~Luca, B{\"u}ergi, Umeton, and
  Caputo}}]{pezziPlasmonmediatedDiscreteDiffraction2019}
\bibinfo{author}{\bibfnamefont{L.}~\bibnamefont{Pezzi}},
  \bibinfo{author}{\bibfnamefont{L.}~\bibnamefont{De~Sio}},
  \bibinfo{author}{\bibfnamefont{A.}~\bibnamefont{Veltri}},
  \bibinfo{author}{\bibfnamefont{A.}~\bibnamefont{Cunningham}},
  \bibinfo{author}{\bibfnamefont{A.}~\bibnamefont{De~Luca}},
  \bibinfo{author}{\bibfnamefont{T.}~\bibnamefont{B{\"u}ergi}},
  \bibinfo{author}{\bibfnamefont{C.}~\bibnamefont{Umeton}}, \bibnamefont{and}
  \bibinfo{author}{\bibfnamefont{R.}~\bibnamefont{Caputo}},
  \bibinfo{journal}{Nanoscale} \textbf{\bibinfo{volume}{11}},
  \bibinfo{pages}{17931} (\bibinfo{year}{2019}).

\bibitem[{\citenamefont{Tran and
  Biancalana}(2013)}]{tranDiffractiveResonantRadiation2013}
\bibinfo{author}{\bibfnamefont{T.~X.} \bibnamefont{Tran}} \bibnamefont{and}
  \bibinfo{author}{\bibfnamefont{F.}~\bibnamefont{Biancalana}},
  \bibinfo{journal}{Phys. Rev. Lett.} \textbf{\bibinfo{volume}{110}},
  \bibinfo{pages}{113903} (\bibinfo{year}{2013}),
  \urlprefix\url{https://link.aps.org/doi/10.1103/PhysRevLett.110.113903}.

\bibitem[{\citenamefont{Karpman}(1993)}]{KarpmanPRE}
\bibinfo{author}{\bibfnamefont{V.~I.} \bibnamefont{Karpman}},
  \bibinfo{journal}{Phys. Rev. E} \textbf{\bibinfo{volume}{47}},
  \bibinfo{pages}{2073} (\bibinfo{year}{1993}),
  \urlprefix\url{https://link.aps.org/doi/10.1103/PhysRevE.47.2073}.

\bibitem[{\citenamefont{Roy et~al.}(2009)\citenamefont{Roy, Bhadra, and
  Agrawal}}]{Roy:09}
\bibinfo{author}{\bibfnamefont{S.}~\bibnamefont{Roy}},
  \bibinfo{author}{\bibfnamefont{S.~K.} \bibnamefont{Bhadra}},
  \bibnamefont{and} \bibinfo{author}{\bibfnamefont{G.~P.}
  \bibnamefont{Agrawal}}, \bibinfo{journal}{Opt. Lett.}
  \textbf{\bibinfo{volume}{34}}, \bibinfo{pages}{2072} (\bibinfo{year}{2009}),
  \urlprefix\url{http://opg.optica.org/ol/abstract.cfm?URI=ol-34-13-2072}.

\bibitem[{\citenamefont{Kr{\'o}likowski and
  Holmstrom}(1997)}]{krolikowskiFusionBirthSpatial1997}
\bibinfo{author}{\bibfnamefont{W.}~\bibnamefont{Kr{\'o}likowski}}
  \bibnamefont{and} \bibinfo{author}{\bibfnamefont{S.~A.}
  \bibnamefont{Holmstrom}}, \bibinfo{journal}{Optics Letters}
  \textbf{\bibinfo{volume}{22}}, \bibinfo{pages}{369} (\bibinfo{year}{1997}).

\bibitem[{\citenamefont{Aossey et~al.}(1992)\citenamefont{Aossey, Skinner,
  Cooney, Williams, Gavin, Andersen, and
  Lonngren}}]{aosseyPropertiesSolitonsolitonCollisions1992}
\bibinfo{author}{\bibfnamefont{D.~W.} \bibnamefont{Aossey}},
  \bibinfo{author}{\bibfnamefont{S.~R.} \bibnamefont{Skinner}},
  \bibinfo{author}{\bibfnamefont{J.~L.} \bibnamefont{Cooney}},
  \bibinfo{author}{\bibfnamefont{J.~E.} \bibnamefont{Williams}},
  \bibinfo{author}{\bibfnamefont{M.~T.} \bibnamefont{Gavin}},
  \bibinfo{author}{\bibfnamefont{D.~R.} \bibnamefont{Andersen}},
  \bibnamefont{and} \bibinfo{author}{\bibfnamefont{K.~E.}
  \bibnamefont{Lonngren}}, \bibinfo{journal}{Physical Review A}
  \textbf{\bibinfo{volume}{45}}, \bibinfo{pages}{2606} (\bibinfo{year}{1992}).

\bibitem[{\citenamefont{Malomed}(1998)}]{malomedPotentialInteractionTwo1998}
\bibinfo{author}{\bibfnamefont{B.~A.} \bibnamefont{Malomed}},
  \bibinfo{journal}{Physical Review E} \textbf{\bibinfo{volume}{58}},
  \bibinfo{pages}{7928} (\bibinfo{year}{1998}).

\bibitem[{\citenamefont{Anastassiou et~al.}(1999)\citenamefont{Anastassiou,
  Segev, Steiglitz, Giordmaine, Mitchell, Shih, Lan, and
  Martin}}]{anastassiouEnergyExchangeInteractionsColliding1999}
\bibinfo{author}{\bibfnamefont{C.}~\bibnamefont{Anastassiou}},
  \bibinfo{author}{\bibfnamefont{M.}~\bibnamefont{Segev}},
  \bibinfo{author}{\bibfnamefont{K.}~\bibnamefont{Steiglitz}},
  \bibinfo{author}{\bibfnamefont{J.~A.} \bibnamefont{Giordmaine}},
  \bibinfo{author}{\bibfnamefont{M.}~\bibnamefont{Mitchell}},
  \bibinfo{author}{\bibfnamefont{M.-f.} \bibnamefont{Shih}},
  \bibinfo{author}{\bibfnamefont{S.}~\bibnamefont{Lan}}, \bibnamefont{and}
  \bibinfo{author}{\bibfnamefont{J.}~\bibnamefont{Martin}},
  \bibinfo{journal}{Physical Review Letters} \textbf{\bibinfo{volume}{83}},
  \bibinfo{pages}{2332} (\bibinfo{year}{1999}).

\bibitem[{\citenamefont{Vahala et~al.}(2004)\citenamefont{Vahala, Vahala, and
  Yepez}}]{vahalaInelasticVectorSoliton2004}
\bibinfo{author}{\bibfnamefont{G.}~\bibnamefont{Vahala}},
  \bibinfo{author}{\bibfnamefont{L.}~\bibnamefont{Vahala}}, \bibnamefont{and}
  \bibinfo{author}{\bibfnamefont{J.}~\bibnamefont{Yepez}},
  \bibinfo{journal}{Philosophical Transactions of the Royal Society of London.
  Series A: Mathematical, Physical and Engineering Sciences}
  \textbf{\bibinfo{volume}{362}}, \bibinfo{pages}{1677} (\bibinfo{year}{2004}).

\bibitem[{\citenamefont{Katsimiga et~al.}(2018)\citenamefont{Katsimiga,
  Kevrekidis, Prinari, Biondini, and
  Schmelcher}}]{katsimigaDarkbrightSolitonPairs2018}
\bibinfo{author}{\bibfnamefont{G.~C.} \bibnamefont{Katsimiga}},
  \bibinfo{author}{\bibfnamefont{P.~G.} \bibnamefont{Kevrekidis}},
  \bibinfo{author}{\bibfnamefont{B.}~\bibnamefont{Prinari}},
  \bibinfo{author}{\bibfnamefont{G.}~\bibnamefont{Biondini}}, \bibnamefont{and}
  \bibinfo{author}{\bibfnamefont{P.}~\bibnamefont{Schmelcher}},
  \bibinfo{journal}{Physical Review A} \textbf{\bibinfo{volume}{97}},
  \bibinfo{pages}{043623} (\bibinfo{year}{2018}).

\bibitem[{\citenamefont{Stalin et~al.}(2021)\citenamefont{Stalin, Ramakrishnan,
  and Lakshmanan}}]{stalinNondegenerateBrightSolitons2021}
\bibinfo{author}{\bibfnamefont{S.}~\bibnamefont{Stalin}},
  \bibinfo{author}{\bibfnamefont{R.}~\bibnamefont{Ramakrishnan}},
  \bibnamefont{and}
  \bibinfo{author}{\bibfnamefont{M.}~\bibnamefont{Lakshmanan}},
  \bibinfo{journal}{Photonics} \textbf{\bibinfo{volume}{8}},
  \bibinfo{pages}{258} (\bibinfo{year}{2021}).

\bibitem[{\citenamefont{Papacharalampous
  et~al.}(2003)\citenamefont{Papacharalampous, Kevrekidis, Malomed, and
  Frantzeskakis}}]{papacharalampousSolitonCollisionsDiscrete2003}
\bibinfo{author}{\bibfnamefont{I.~E.} \bibnamefont{Papacharalampous}},
  \bibinfo{author}{\bibfnamefont{P.~G.} \bibnamefont{Kevrekidis}},
  \bibinfo{author}{\bibfnamefont{B.~A.} \bibnamefont{Malomed}},
  \bibnamefont{and} \bibinfo{author}{\bibfnamefont{D.~J.}
  \bibnamefont{Frantzeskakis}}, \bibinfo{journal}{Physical Review E}
  \textbf{\bibinfo{volume}{68}}, \bibinfo{pages}{046604}
  (\bibinfo{year}{2003}).

\bibitem[{\citenamefont{Al~Khawaja et~al.}(2016)\citenamefont{Al~Khawaja,
  {Al-Marzoug}, Bahlouli, and
  Baizakov}}]{alkhawajaInteractionPotentialDiscrete2016}
\bibinfo{author}{\bibfnamefont{U.}~\bibnamefont{Al~Khawaja}},
  \bibinfo{author}{\bibfnamefont{S.~M.} \bibnamefont{{Al-Marzoug}}},
  \bibinfo{author}{\bibfnamefont{H.}~\bibnamefont{Bahlouli}}, \bibnamefont{and}
  \bibinfo{author}{\bibfnamefont{B.}~\bibnamefont{Baizakov}},
  \bibinfo{journal}{Optics Express} \textbf{\bibinfo{volume}{24}},
  \bibinfo{pages}{18148} (\bibinfo{year}{2016}).

\bibitem[{\citenamefont{Xiao et~al.}(2011)\citenamefont{Xiao, Zhang, Liu, and
  Zhao}}]{xiaoTunableOscillationDiscrete2011}
\bibinfo{author}{\bibfnamefont{F.}~\bibnamefont{Xiao}},
  \bibinfo{author}{\bibfnamefont{P.}~\bibnamefont{Zhang}},
  \bibinfo{author}{\bibfnamefont{S.}~\bibnamefont{Liu}}, \bibnamefont{and}
  \bibinfo{author}{\bibfnamefont{J.}~\bibnamefont{Zhao}},
  \bibinfo{journal}{Journal of Optics} \textbf{\bibinfo{volume}{13}},
  \bibinfo{pages}{105101} (\bibinfo{year}{2011}).

\bibitem[{\citenamefont{Cuevas and
  Eilbeck}(2006)}]{cuevasDiscreteSolitonCollisions2006}
\bibinfo{author}{\bibfnamefont{J.}~\bibnamefont{Cuevas}} \bibnamefont{and}
  \bibinfo{author}{\bibfnamefont{J.}~\bibnamefont{Eilbeck}},
  \bibinfo{journal}{Physics Letters A} \textbf{\bibinfo{volume}{358}},
  \bibinfo{pages}{15} (\bibinfo{year}{2006}), \eprint{nlin/0501050}.

\bibitem[{\citenamefont{Parasuraman}(2019)}]{parasuramanDynamicsSolitonCollision2019a}
\bibinfo{author}{\bibfnamefont{E.}~\bibnamefont{Parasuraman}},
  \bibinfo{journal}{Journal of Magnetism and Magnetic Materials}
  \textbf{\bibinfo{volume}{489}}, \bibinfo{pages}{165403}
  (\bibinfo{year}{2019}).

\bibitem[{\citenamefont{Gao et~al.}(2021)\citenamefont{Gao, Li, Yang, Yang, and
  Yi}}]{gaoBreathingSolitonsInduced2021a}
\bibinfo{author}{\bibfnamefont{P.}~\bibnamefont{Gao}},
  \bibinfo{author}{\bibfnamefont{X.}~\bibnamefont{Li}},
  \bibinfo{author}{\bibfnamefont{Z.-Y.} \bibnamefont{Yang}},
  \bibinfo{author}{\bibfnamefont{W.-L.} \bibnamefont{Yang}}, \bibnamefont{and}
  \bibinfo{author}{\bibfnamefont{S.}~\bibnamefont{Yi}},
  \textbf{\bibinfo{volume}{54}}, \bibinfo{pages}{135301}
  (\bibinfo{year}{2021}).

\bibitem[{\citenamefont{Christov et~al.}(1994)\citenamefont{Christov, Dost, and
  Maugin}}]{christovInelasticitySolitonCollisions1994}
\bibinfo{author}{\bibfnamefont{C.~I.} \bibnamefont{Christov}},
  \bibinfo{author}{\bibfnamefont{S.}~\bibnamefont{Dost}}, \bibnamefont{and}
  \bibinfo{author}{\bibfnamefont{G.~A.} \bibnamefont{Maugin}},
  \bibinfo{journal}{Physica Scripta} \textbf{\bibinfo{volume}{50}},
  \bibinfo{pages}{449} (\bibinfo{year}{1994}).

\bibitem[{\citenamefont{Solja{\u c}i{\'c} et~al.}(2003)\citenamefont{Solja{\u
  c}i{\'c}, Steiglitz, Sears, Segev, Jakubowski, and
  Squier}}]{soljacicCollisionsTwoSolitons2003}
\bibinfo{author}{\bibfnamefont{M.}~\bibnamefont{Solja{\u c}i{\'c}}},
  \bibinfo{author}{\bibfnamefont{K.}~\bibnamefont{Steiglitz}},
  \bibinfo{author}{\bibfnamefont{S.~M.} \bibnamefont{Sears}},
  \bibinfo{author}{\bibfnamefont{M.}~\bibnamefont{Segev}},
  \bibinfo{author}{\bibfnamefont{M.~H.} \bibnamefont{Jakubowski}},
  \bibnamefont{and} \bibinfo{author}{\bibfnamefont{R.}~\bibnamefont{Squier}},
  \bibinfo{journal}{Physical Review Letters} \textbf{\bibinfo{volume}{90}},
  \bibinfo{pages}{254102} (\bibinfo{year}{2003}).

\bibitem[{\citenamefont{Anderson}(1983)}]{PhysRevA.27.3135}
\bibinfo{author}{\bibfnamefont{D.}~\bibnamefont{Anderson}},
  \bibinfo{journal}{Phys. Rev. A} \textbf{\bibinfo{volume}{27}},
  \bibinfo{pages}{3135} (\bibinfo{year}{1983}),
  \urlprefix\url{https://link.aps.org/doi/10.1103/PhysRevA.27.3135}.

\bibitem[{\citenamefont{Lederer and
  Silberberg}(2002)}]{ledererDiscreteSolitons2002}
\bibinfo{author}{\bibfnamefont{F.}~\bibnamefont{Lederer}} \bibnamefont{and}
  \bibinfo{author}{\bibfnamefont{Y.}~\bibnamefont{Silberberg}},
  \bibinfo{journal}{Optics and Photonics News} \textbf{\bibinfo{volume}{13}},
  \bibinfo{pages}{48} (\bibinfo{year}{2002}).

\bibitem[{\citenamefont{Eisenberg et~al.}(2000)\citenamefont{Eisenberg,
  Silberberg, Morandotti, and Aitchison}}]{eisenbergDiffractionManagement2000}
\bibinfo{author}{\bibfnamefont{H.~S.} \bibnamefont{Eisenberg}},
  \bibinfo{author}{\bibfnamefont{Y.}~\bibnamefont{Silberberg}},
  \bibinfo{author}{\bibfnamefont{R.}~\bibnamefont{Morandotti}},
  \bibnamefont{and} \bibinfo{author}{\bibfnamefont{J.~S.}
  \bibnamefont{Aitchison}}, \bibinfo{journal}{Physical Review Letters}
  \textbf{\bibinfo{volume}{85}}, \bibinfo{pages}{1863} (\bibinfo{year}{2000}).

\bibitem[{\citenamefont{Szameit et~al.}(2006)\citenamefont{Szameit, Bl{\"o}mer,
  Burghoff, Pertsch, Nolte, and
  T{\"u}nnermann}}]{szameitHexagonalWaveguideArrays2006a}
\bibinfo{author}{\bibfnamefont{A.}~\bibnamefont{Szameit}},
  \bibinfo{author}{\bibfnamefont{D.}~\bibnamefont{Bl{\"o}mer}},
  \bibinfo{author}{\bibfnamefont{J.}~\bibnamefont{Burghoff}},
  \bibinfo{author}{\bibfnamefont{T.}~\bibnamefont{Pertsch}},
  \bibinfo{author}{\bibfnamefont{S.}~\bibnamefont{Nolte}}, \bibnamefont{and}
  \bibinfo{author}{\bibfnamefont{A.}~\bibnamefont{T{\"u}nnermann}},
  \bibinfo{journal}{Applied Physics B} \textbf{\bibinfo{volume}{82}},
  \bibinfo{pages}{507} (\bibinfo{year}{2006}).

\bibitem[{\citenamefont{Pavlov et~al.}(2017)\citenamefont{Pavlov, Tokel,
  Pavlova, Kadan, Makey, Turnali, Yavuz, and
  Ilday}}]{pavlovFemtosecondLaserWritten2017}
\bibinfo{author}{\bibfnamefont{I.}~\bibnamefont{Pavlov}},
  \bibinfo{author}{\bibfnamefont{O.}~\bibnamefont{Tokel}},
  \bibinfo{author}{\bibfnamefont{S.}~\bibnamefont{Pavlova}},
  \bibinfo{author}{\bibfnamefont{V.}~\bibnamefont{Kadan}},
  \bibinfo{author}{\bibfnamefont{G.}~\bibnamefont{Makey}},
  \bibinfo{author}{\bibfnamefont{A.}~\bibnamefont{Turnali}},
  \bibinfo{author}{\bibfnamefont{{\"O}.}~\bibnamefont{Yavuz}},
  \bibnamefont{and} \bibinfo{author}{\bibfnamefont{F.~{\"O}.}
  \bibnamefont{Ilday}}, \bibinfo{journal}{Optics Letters}
  \textbf{\bibinfo{volume}{42}}, \bibinfo{pages}{3028} (\bibinfo{year}{2017}).

\bibitem[{\citenamefont{Tewari and
  Thyagarajan}(1986)}]{tewariAnalysisTunableSinglemode1986}
\bibinfo{author}{\bibfnamefont{R.}~\bibnamefont{Tewari}} \bibnamefont{and}
  \bibinfo{author}{\bibfnamefont{K.}~\bibnamefont{Thyagarajan}},
  \bibinfo{journal}{Journal of Lightwave Technology}
  \textbf{\bibinfo{volume}{4}}, \bibinfo{pages}{386} (\bibinfo{year}{1986}).

\bibitem[{\citenamefont{Snyder}(1972)}]{snyderCoupledModeTheoryOptical1972}
\bibinfo{author}{\bibfnamefont{A.~W.} \bibnamefont{Snyder}},
  \bibinfo{journal}{Journal of the Optical Society of America}
  \textbf{\bibinfo{volume}{62}}, \bibinfo{pages}{1267} (\bibinfo{year}{1972}).

\bibitem[{\citenamefont{Sakai et~al.}(1990)\citenamefont{Sakai, Hawkins, and
  Friberg}}]{sakaiSolitoncollisionInterferometerQuantum1990}
\bibinfo{author}{\bibfnamefont{Y.}~\bibnamefont{Sakai}},
  \bibinfo{author}{\bibfnamefont{R.~J.} \bibnamefont{Hawkins}},
  \bibnamefont{and} \bibinfo{author}{\bibfnamefont{S.~R.}
  \bibnamefont{Friberg}}, \bibinfo{journal}{Optics Letters}
  \textbf{\bibinfo{volume}{15}}, \bibinfo{pages}{239} (\bibinfo{year}{1990}).

\bibitem[{mov({\natexlab{a}})}]{movie1}
\emph{\bibinfo{title}{See supplemental material at [url will be inserted by
  publisher] for a spatial spectrogram-movie describing the evolution of the
  elastic collision of two identical solitons.}}, \urlprefix\url{url}.

\bibitem[{mov({\natexlab{b}})}]{movie2}
\emph{\bibinfo{title}{See supplemental material at [url will be inserted by
  publisher] for a spatial spectrogram-movie describing the evolution of
  breather formation from the collision of two identical solitons.}},
  \urlprefix\url{url}.

\bibitem[{\citenamefont{Blow et~al.}(1992)\citenamefont{Blow, Doran, and
  Phoenix}}]{blowSolitonPhase1992a}
\bibinfo{author}{\bibfnamefont{K.}~\bibnamefont{Blow}},
  \bibinfo{author}{\bibfnamefont{N.}~\bibnamefont{Doran}}, \bibnamefont{and}
  \bibinfo{author}{\bibfnamefont{S.}~\bibnamefont{Phoenix}},
  \bibinfo{journal}{Optics Communications} \textbf{\bibinfo{volume}{88}},
  \bibinfo{pages}{137} (\bibinfo{year}{1992}).

\bibitem[{\citenamefont{Stegeman and
  Segev}(1999)}]{stegemanOpticalSpatialSolitons1999a}
\bibinfo{author}{\bibfnamefont{G.~I.} \bibnamefont{Stegeman}} \bibnamefont{and}
  \bibinfo{author}{\bibfnamefont{M.}~\bibnamefont{Segev}},
  \bibinfo{journal}{Science} \textbf{\bibinfo{volume}{286}},
  \bibinfo{pages}{1518} (\bibinfo{year}{1999}),
  \eprint{https://www.science.org/doi/pdf/10.1126/science.286.5444.1518},
  \urlprefix\url{https://www.science.org/doi/abs/10.1126/science.286.5444.1518}.

\bibitem[{mov({\natexlab{c}})}]{movie3}
\emph{\bibinfo{title}{See supplemental material at [url will be inserted by
  publisher] for a spatial spectrogram evolution movie for collision of two
  nonidentical solitons resulting in the formation of collision mediated
  radiation.}}, \urlprefix\url{url}.

\bibitem[{\citenamefont{Tran and
  {Nguyen-The}}(2016)}]{tranControllingDiscreteSoliton2016}
\bibinfo{author}{\bibfnamefont{T.~X.} \bibnamefont{Tran}} \bibnamefont{and}
  \bibinfo{author}{\bibfnamefont{Q.}~\bibnamefont{{Nguyen-The}}},
  \bibinfo{journal}{Journal of Lightwave Technology}
  \textbf{\bibinfo{volume}{34}}, \bibinfo{pages}{4105} (\bibinfo{year}{2016}).

\bibitem[{\citenamefont{Meier et~al.}(2005)\citenamefont{Meier, Stegeman,
  Christodoulides, Silberberg, Morandotti, Yang, Salamo, Sorel, and
  Aitchison}}]{Meier:05}
\bibinfo{author}{\bibfnamefont{J.}~\bibnamefont{Meier}},
  \bibinfo{author}{\bibfnamefont{G.~I.} \bibnamefont{Stegeman}},
  \bibinfo{author}{\bibfnamefont{D.~N.} \bibnamefont{Christodoulides}},
  \bibinfo{author}{\bibfnamefont{Y.}~\bibnamefont{Silberberg}},
  \bibinfo{author}{\bibfnamefont{R.}~\bibnamefont{Morandotti}},
  \bibinfo{author}{\bibfnamefont{H.}~\bibnamefont{Yang}},
  \bibinfo{author}{\bibfnamefont{G.}~\bibnamefont{Salamo}},
  \bibinfo{author}{\bibfnamefont{M.}~\bibnamefont{Sorel}}, \bibnamefont{and}
  \bibinfo{author}{\bibfnamefont{J.~S.} \bibnamefont{Aitchison}},
  \bibinfo{journal}{Opt. Lett.} \textbf{\bibinfo{volume}{30}},
  \bibinfo{pages}{1027} (\bibinfo{year}{2005}),
  \urlprefix\url{http://opg.optica.org/ol/abstract.cfm?URI=ol-30-9-1027}.

\end{thebibliography}
	\bibliographystyle{apsrev}
\end{document}